\documentclass[zpreprint,zbstepj]{./LaTeX/zeus/zeus_paper}
\usepackage[english]{babel}

\newcommand{\ZcoosysB}{%
The ZEUS coordinate system is a right-handed Cartesian system, with the $Z$
axis pointing in the proton beam direction, referred to as the ``forward
direction'', and the $X$ axis pointing left towards the centre of HERA.
The coordinate origin is at the nominal interaction point.\xspace}
\newcommand{\Zpsrap}{%
The pseudorapidity is defined as $\eta=-\ln\left(\tan\frac{\theta}{2}\right)$,
where the polar angle, $\theta$, is measured with respect to the proton beam
direction.\xspace}

\newcommand{\ZcoosysfnBeta}{\footnote{\ZcoosysB\Zpsrap}}

\newcommand{\Zctddesc}[1]{%
Charged particles are tracked in the central tracking detector (CTD)~\citeCTD,
which operates in a magnetic field of $1.43\Tesla$ provided by a thin 
superconducting coil. The CTD consists of 72~cylindrical drift chamber 
layers, organized in 9~superlayers covering the polar angle#1 region 
\mbox{$15^\circ<\theta<164^\circ$}. The transverse-momentum resolution for
full-length tracks is $\sigma(p_T)/p_T=0.0058p_T\oplus0.0065\oplus0.0014/p_T$,
with $p_T$ in $\Gev$.}
\newcommand{\Zcaldesc}{%
The high-resolution uranium--scintillator calorimeter (CAL)~\citeCAL consists 
of three parts: the forward (FCAL), the barrel (BCAL) and the rear (RCAL)
calorimeters. Each part is subdivided transversely into towers and
longitudinally into one electromagnetic section (EMC) and either one (in RCAL)
or two (in BCAL and FCAL) hadronic sections (HAC). The smallest subdivision of
the calorimeter is called a cell.  The CAL energy resolutions, as measured under
test beam conditions, are $\sigma(E)/E=0.18/\sqrt{E}$ for electrons and
$\sigma(E)/E=0.35/\sqrt{E}$ for hadrons ($E$ in $\Gev$).}

\chardef\usc=95
\chardef\til=126
\catcode`\@=11 
\DeclareRobustCommand\xdotspace{\futurelet\@let@token\@xdotspace}
\def\@xdotspace{%
  \ifx\@let@token.\else
  \ifx\@let@token\bgroup.\else
  \ifx\@let@token\egroup.\else
  \ifx\@let@token\/.\else
  \ifx\@let@token\ .\else
  \ifx\@let@token~.\else
  \ifx\@let@token!.\else
  \ifx\@let@token,.\else
  \ifx\@let@token:.\else
  \ifx\@let@token;.\else
  \ifx\@let@token?.\else
  \ifx\@let@token/.\else
  \ifx\@let@token'.\else
  \ifx\@let@token).\else
  \ifx\@let@token-.\else
  \ifx\@let@token\@xobeysp.\else
  \ifx\@let@token\space.\else
  \ifx\@let@token\@sptoken.\else
   .\space
   \fi\fi\fi\fi\fi\fi\fi\fi\fi\fi\fi\fi\fi\fi\fi\fi\fi\fi}
\catcode`\@=12 

\newcommand{\stru}[2]{%
   \relax\ifmmode\hbox{\vrule height#1 depth#2 width0pt}%
   \else\vrule height#1 depth#2 width0pt\fi}

\newcommand{\Ronum}[1]{\uppercase\expandafter{\romannumeral#1}}
\newcommand{\ronum}[1]{\expandafter{\romannumeral#1}}
\DeclareRobustCommand{\LaTeXZ}{%
  \LaTeX\kern-.05em4\kern-.1em
  {\raisebox{-0.2ex}{$\scriptstyle\text{ZEUS}$}}\xspace}

\newcommand{\eq}[1]{(\ref{eq-#1})}

\newcommand{\fig}[1]{Fig.~\ref{fig-#1}}

\DeclareMathAlphabet{\mathbf}{OT1}{cmr}{bx}{sl}
\newcommand{\eVdist}{\kern-0.06667em}

\newcommand{\Gev}{{\text{Ge}\eVdist\text{V\/}}}

\newcommand{\Tesla}{\,\text{T}}

\newcommand{\slashfrac}[2]{%
  \raisebox{0.5ex}{\ensuremath #1}\kern-0.12em/\kern-0.08em
  \raisebox{-.8ex}{\ensuremath #2}}

\newcommand{\sqr}[3]{%
    {\vcenter{\hrule height.#3ex\hbox{\vrule width.#2ex height#1ex
     \kern#1ex\vrule width.#3ex}\hrule height.#2ex}}}

\catcode`\@=11 
\newcommand{\parenbar}{\mathpalette\p@renb@r}
\def\p@renb@r#1#2{\vbox{%
  \ifx#1\scriptscriptstyle \dimen@.7em\dimen@ii.2em\else
  \ifx#1\scriptstyle \dimen@.8em\dimen@ii.25em\else
  \dimen@1em\dimen@ii.4em\fi\fi \offinterlineskip
  \ialign{\hfill##\hfill\cr
    \vbox{\hrule width\dimen@ii}\cr
    \noalign{\vskip-.3ex}%
    \hbox to\dimen@{$\mathchar300\hfil\mathchar301$}\cr
    \noalign{\vskip-.3ex}%
    $#1#2$\cr}}}
\catcode`\@=12 

\newcommand{\IP}{{\rm I$\kern-0.01667em$P}\xspace}

\mathchardef\qsm=63
\mathchardef\pls=43
\mathchardef\mns=512
\mathchardef\plm=518
\mathchardef\eql=61
\mathchardef\smallleft=300
\mathchardef\smallright=301
\mathchardef\les=316
\mathchardef\gre=318
\mathchardef\leq=532
\mathchardef\grq=533

\catcode`\@=11 
\newcounter{pict@width}
\newcounter{pict@height}
\newlength{\pict@scale}
\newcommand{\psfigadd}[4]{%
\setcounter{pict@width}{1*\ratio{#2+\pict@scale/2}{\pict@scale}}
\setcounter{pict@height}{1*\ratio{#3+\pict@scale/2}{\pict@scale}}
\setlength{\unitlength}{\pict@scale}
\hbox to #2{\hspace{-\fill}\begin{picture}(\thepict@width,\thepict@height)
\put(0,0){\psfig{figure=#1,width=#2,height=#3,clip=}}
\SetScale{0.283466457}
\SetWidth{1.763889}
{#4}
\end{picture}}
}
\newcounter{pict@widthfst}
\newcounter{pict@widthscd}
\newcounter{pict@widthtot}
\newcommand{\psfigaddtwo}[7]{%
\setcounter{pict@widthfst}{1*\ratio{#2+\pict@scale/2}{\pict@scale}}
\setcounter{pict@widthscd}{1*\ratio{#2+#4+\pict@scale/2}{\pict@scale}}
\setcounter{pict@widthtot}{1*\ratio{#2+#4+#6+\pict@scale/2}{\pict@scale}}
\setcounter{pict@height}{1*\ratio{#3+\pict@scale/2}{\pict@scale}}
\setlength{\unitlength}{\pict@scale}
\hbox{\hspace{-\fill}\begin{picture}(\thepict@widthtot,\thepict@height)
\put(0,0){\psfig{figure=#1,width=#2,height=#3,clip=}}
\put(\thepict@widthscd,0){\psfig{figure=#5,width=#6,height=#3,clip=}}
\SetScale{0.283466457}
\SetWidth{1.763889}
{#7}
\end{picture}}
}
\newcommand{\psfigror}[4]{%
\setcounter{pict@width}{1*\ratio{#2+\pict@scale/2}{\pict@scale}}
\setcounter{pict@height}{1*\ratio{#3+\pict@scale/2}{\pict@scale}}
\setlength{\unitlength}{\pict@scale}
\hbox{\begin{picture}(\thepict@width,\thepict@height)
\put(0,\thepict@height){\psfig{figure=#1,width=#3,height=#2,clip=,angle=270}}
\SetScale{0.283466457}
\SetWidth{1.763889}
{#4}
\end{picture}}
}
\newcommand{\psfigrol}[4]{%
\setcounter{pict@width}{1*\ratio{#2+\pict@scale/2}{\pict@scale}}
\setcounter{pict@height}{1*\ratio{#3+\pict@scale/2}{\pict@scale}}
\setlength{\unitlength}{\pict@scale}
\hbox{\begin{picture}(\thepict@width,\thepict@height)
\put(0,0){\psfig{figure=#1,width=#3,height=#2,clip=,angle=90}}
\SetScale{0.283466457}
\SetWidth{1.763889}
{#4}
\end{picture}}
}
\catcode`\@=12 
\newlength\listtextwidth

\catcode`\@=11 
\newlength{\@tabfninsert}
\newlength{\@tabfnwidth}
\newcommand{\tabfootnote}[2]{%
  \setlength{\@tabfninsert}{0.8em}
  \setlength{\@tabfnwidth}{\textwidth}
  \addtolength{\@tabfnwidth}{-\@tabfninsert}
  \addtolength{\@tabfnwidth}{-0.4em}
  \noindent\makebox[\@tabfninsert][r]{\footnotesize$^{#1}$\hfil}\hfill%
  \parbox[t]{\@tabfnwidth}{\footnotesize #2\hfill}}
\catcode`\@=12 

\def\citeCTD{{\cite{%
nim:a279:290,*npps:b32:181,*nim:a338:254%
}}\xspace}
\def\citeCAL{{\cite{%
nim:a309:77,*nim:a309:101,*nim:a321:356,*nim:a336:23%
}}\xspace}

\begin{document}

\prepnum{DESY--02--163\\
October 2002} 

\title{
Measurements of inelastic $J/\psi$ and $\psi^{\prime}$ photoproduction at HERA
}

\author{ZEUS Collaboration} 
\draftversion{post reading}

\date{}

\abstract{
The cross sections for inelastic photoproduction of $J/\psi$ and $\psi^{\prime}$ mesons 
have been measured in $ep$ collisions with the ZEUS detector at HERA, using an 
integrated luminosity of 38.0 pb$^{-1}$. The events were required to have 
\mbox{0.1 $< z <$ 0.9} and \mbox{50 $< W <$ 180 GeV}, where $z$ is the fraction of the 
incident photon energy carried by the $J/\psi$ in the proton rest frame and $W$ is the 
photon-proton centre-of-mass energy. The $\psi^{\prime}$ to $J/\psi$ cross-section ratio 
was measured in the range \mbox{$0.55 < z < 0.9$}. The $J/\psi$ data, for various ranges 
of transverse momentum, are compared to theoretical models incorporating colour-singlet 
and colour-octet matrix elements. Predictions of a next-to-leading-order colour-singlet 
model give a good description of the data, although there is a large normalisation 
uncertainty. The $J/\psi$ helicity distribution for $z >$~0.4 is compared to 
leading-order QCD predictions.
}

\makezeustitle

\def\3{\ss}                                                                                        
\newcommand{\address}{ }                                                                           
\pagenumbering{Roman}                                                                              
                                                   %
\begin{center}                                                                                     
{                      \Large  The ZEUS Collaboration              }                               
\end{center}                                                                                       
  S.~Chekanov,                                                                                     
  D.~Krakauer,                                                                                     
  S.~Magill,                                                                                       
  B.~Musgrave,                                                                                     
  J.~Repond,                                                                                       
  R.~Yoshida\\                                                                                     
 {\it Argonne National Laboratory, Argonne, Illinois 60439-4815}~$^{n}$                            
\par \filbreak                                                                                     
  M.C.K.~Mattingly \\                                                                              
 {\it Andrews University, Berrien Springs, Michigan 49104-0380}                                    
\par \filbreak                                                                                     
  P.~Antonioli,                                                                                    
  G.~Bari,                                                                                         
  M.~Basile,                                                                                       
  L.~Bellagamba,                                                                                   
  D.~Boscherini,                                                                                   
  A.~Bruni,                                                                                        
  G.~Bruni,                                                                                        
  G.~Cara~Romeo,                                                                                   
  L.~Cifarelli,                                                                                    
  F.~Cindolo,                                                                                      
  A.~Contin,                                                                                       
  M.~Corradi,                                                                                      
  S.~De~Pasquale,                                                                                  
  P.~Giusti,                                                                                       
  G.~Iacobucci,                                                                                    
  A.~Margotti,                                                                                     
  R.~Nania,                                                                                        
  F.~Palmonari,                                                                                    
  A.~Pesci,                                                                                        
  G.~Sartorelli,                                                                                   
  A.~Zichichi  \\                                                                                  
  {\it University and INFN Bologna, Bologna, Italy}~$^{e}$
% ...
\newpage                                                                                           
$^{\    1}$ on leave of absence at University of                                                   
Erlangen-N\"urnberg, Germany\\                                                                     
% ...
$^{\    9}$ member of Dept. of Computer Science \\                                                 
% ...
\par         
                                                           %
                                                           %
\begin{tabular}[h]{rp{14cm}}                                                                       
$^{a}$ &  supported by the Natural Sciences and Engineering Research                               
          Council of Canada (NSERC) \\                                                             
$^{b}$ &  supported by the German Federal Ministry for Education and                               
          Research (BMBF), under contract numbers HZ1GUA 2, HZ1GUB 0, HZ1PDA 5, HZ1VFA 5\\         
$^{c}$ &  supported by the MINERVA Gesellschaft f\"ur Forschung GmbH, the                          
          Israel Science Foundation, the U.S.-Israel Binational Science                            
          Foundation and the Benozyio Center                                                       
          for High Energy Physics\\                                                                
$^{d}$ &  supported by the German-Israeli Foundation and the Israel Science                        
          Foundation\\                                                                             
$^{e}$ &  supported by the Italian National Institute for Nuclear Physics (INFN) \\                
$^{f}$ &  supported by the Japanese Ministry of Education, Science and                             
          Culture (the Monbusho) and its grants for Scientific Research\\                          
$^{g}$ &  supported by the Korean Ministry of Education and Korea Science                          
          and Engineering Foundation\\                                                             
$^{h}$ &  supported by the Netherlands Foundation for Research on Matter (FOM)\\                   
$^{i}$ &  supported by the Polish State Committee for Scientific Research,                         
          grant no. 620/E-77/SPUB-M/DESY/P-03/DZ 247/2000-2002\\                                   
$^{j}$ &  partially supported by the German Federal Ministry for Education                         
          and Research (BMBF)\\                                                                    
$^{k}$ &  supported by the Fund for Fundamental Research of Russian Ministry                       
          for Science and Edu\-cation and by the German Federal Ministry for                       
          Education and Research (BMBF)\\                                                          
$^{l}$ &  supported by the Spanish Ministry of Education and Science                               
          through funds provided by CICYT\\                                                        
$^{m}$ &  supported by the Particle Physics and Astronomy Research Council, UK\\                   
$^{n}$ &  supported by the US Department of Energy\\                                               
$^{o}$ &  supported by the US National Science Foundation\\                                        
$^{p}$ &  supported by the Polish State Committee for Scientific Research,                         
          grant no. 112/E-356/SPUB-M/DESY/P-03/DZ 301/2000-2002, 2 P03B 13922\\                    
$^{q}$ &  supported by the Polish State Committee for Scientific Research,                         
          grant no. 115/E-343/SPUB-M/DESY/P-03/DZ 121/2001-2002, 2 P03B 07022\\                    
\end{tabular}                                                                                      
                                                           %
                                                           %

\pagenumbering{arabic} 
\pagestyle{plain}
\section{Introduction}
\label{sec:intro}

In the HERA photoproduction regime, where the virtuality of the exchanged photon 
is small, the production of inelastic $\psi$ mesons, where $\psi$ can be either a 
$J/\psi$ or a $\psi^{\prime}$, arises mostly from direct and resolved photon 
interactions. In leading-order (LO) Quantum Chromodynamics (QCD), the two 
processes can be distinguished: in direct photon processes, the photon couples 
directly to a parton in the proton; in resolved photon processes, the photon acts 
as a source of partons, one of which participates in the hard interaction. 
Diffractive production, $\gamma p \rightarrow \psi N$, where $N$ is a 
proton-dissociative state, contributes significantly to the inelastic production 
of $\psi$ mesons by the direct photon process.

\begin{figure}
\unitlength1cm  \begin{picture}(15.,15.)
\includegraphics{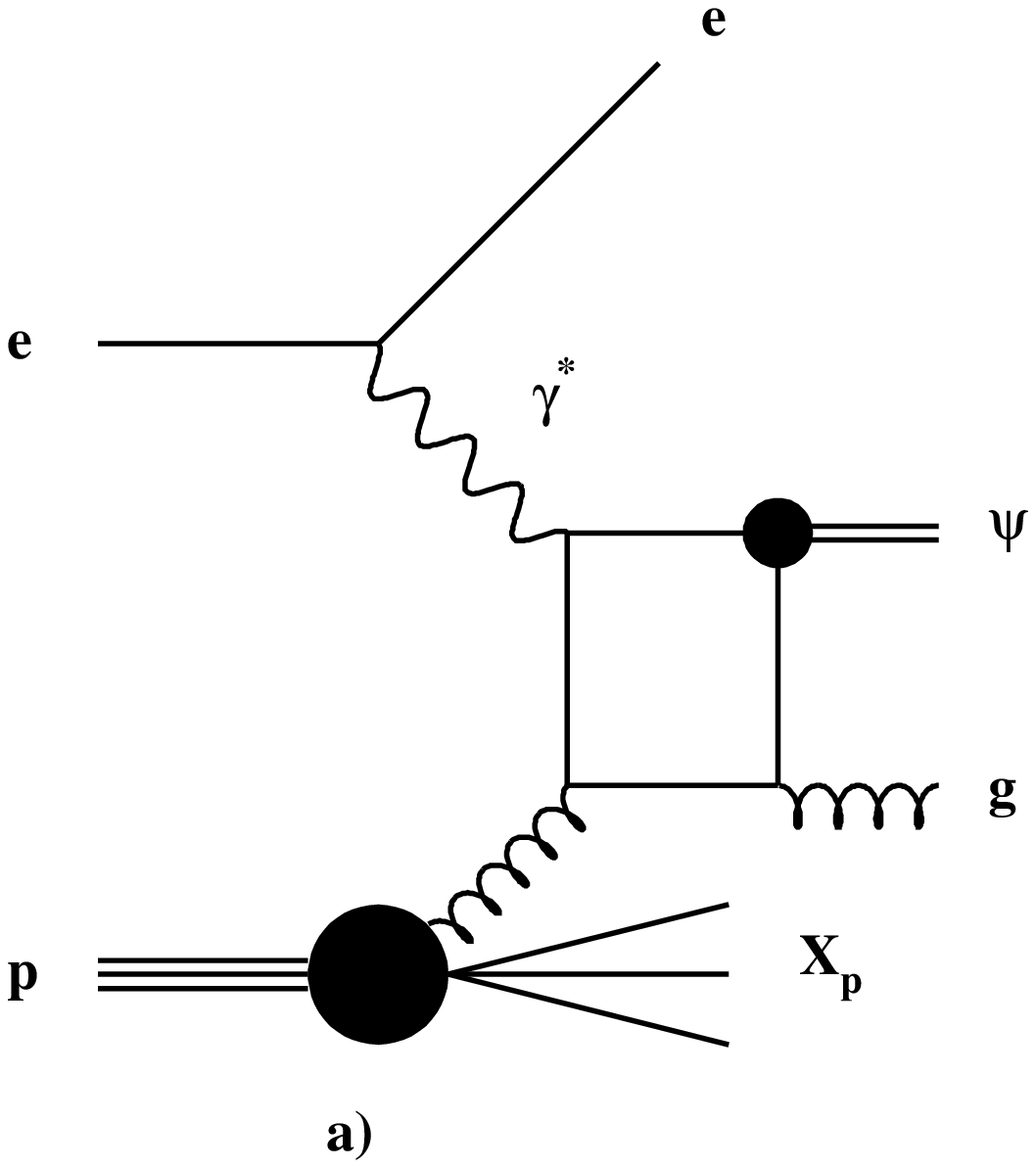}
\includegraphics{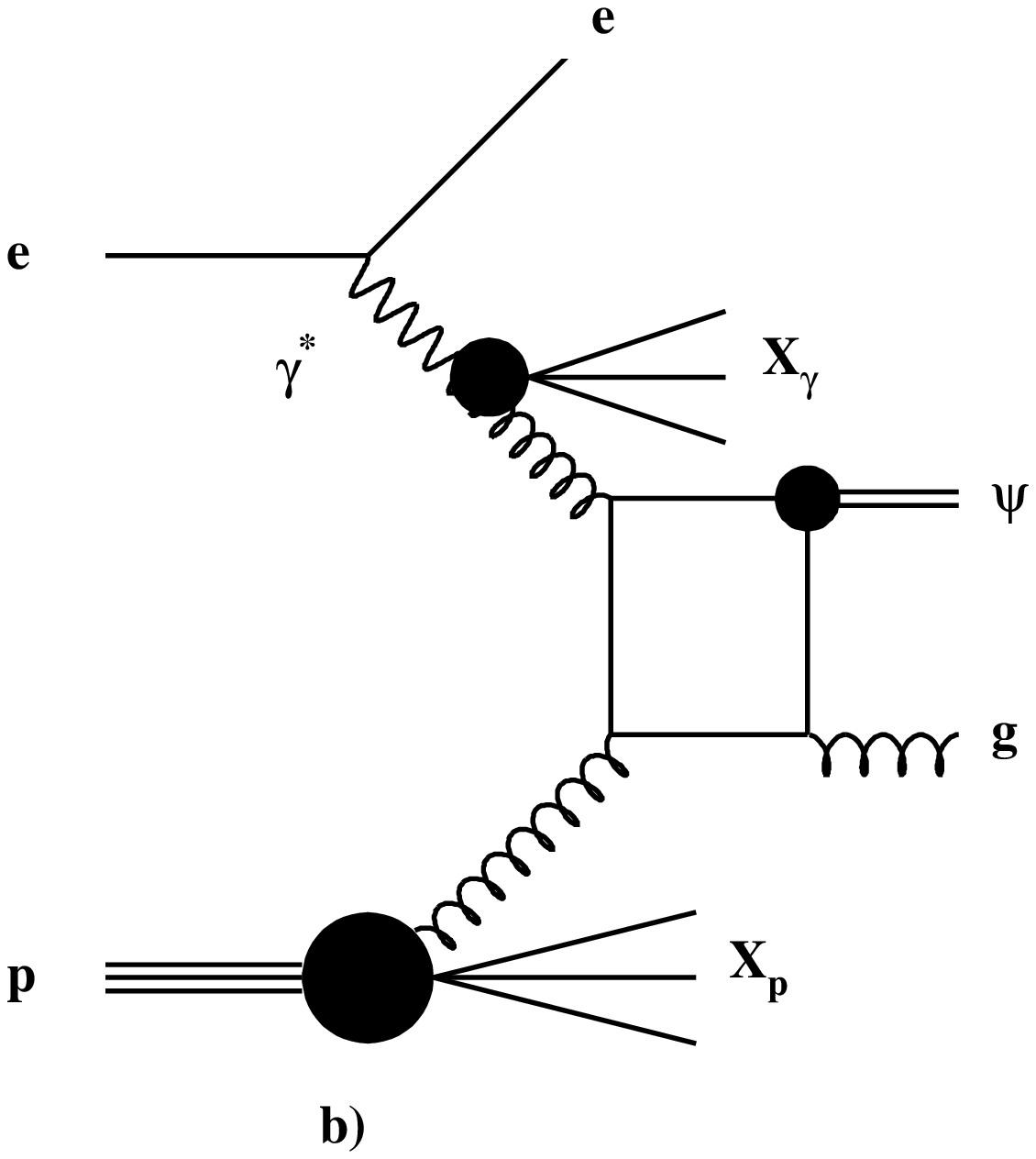}
\includegraphics{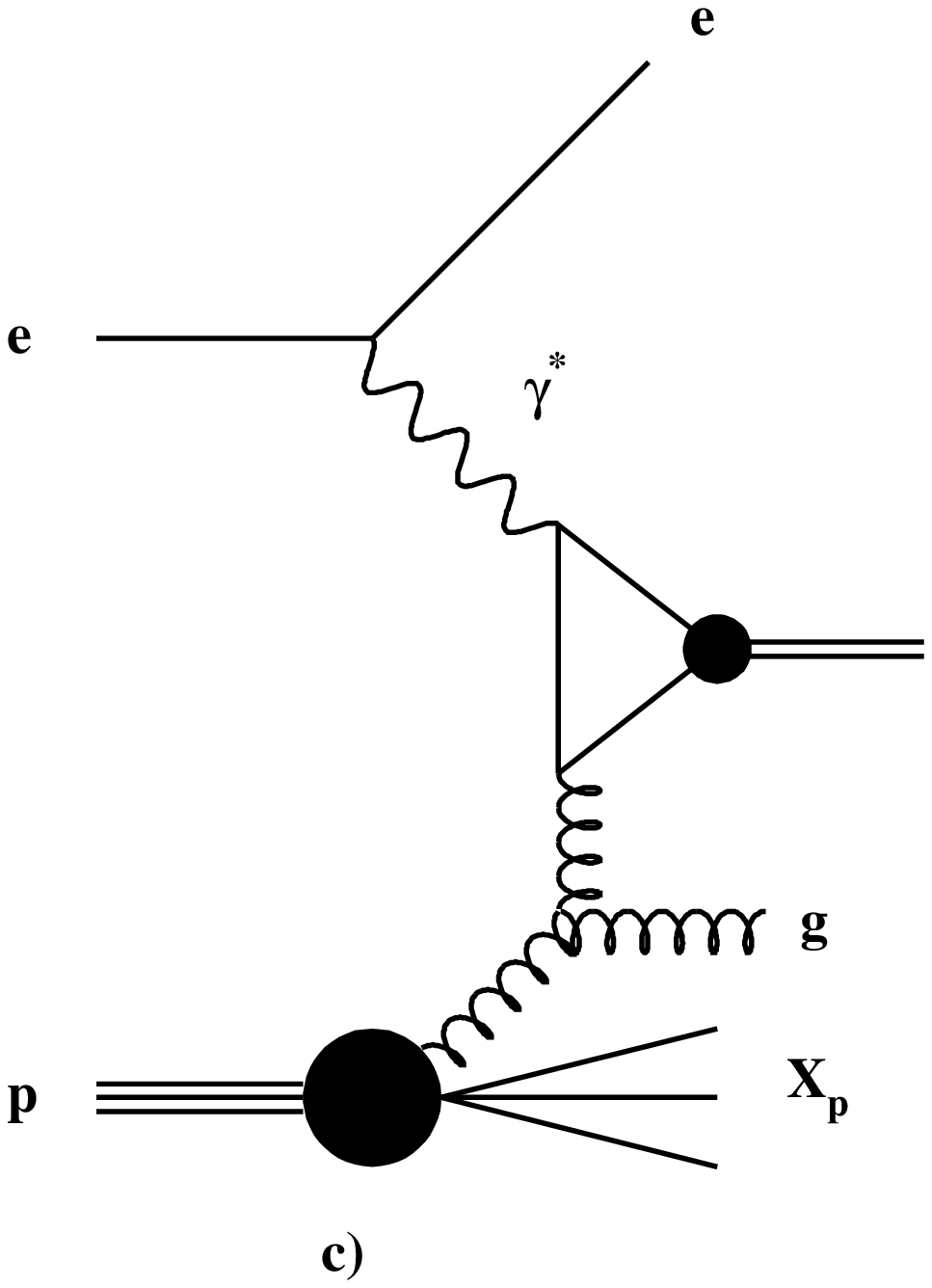}
\end{picture}
\caption{a) The direct photon process at leading order in the CS 
framework; b) the resolved photon process in the same framework; c) 
the direct photon process in the CO framework.}
\label{fig-fey}
\end{figure}

Direct and resolved photon cross sections can be calculated using perturbative QCD 
(pQCD) in the colour-singlet (CS) and colour-octet (CO) frameworks~\cite{pr:d23:1521,
pl:b348:657,*np:b459:3,hep-ph-0106120,prl:76:4128,*pr:d54:4312,*pr:d60:119902,
pr:d62:34004,epj:c6:493,pr:d57:4258,pl:b428:377}. In the CS model, the colourless 
$c \bar{c}$ pair produced by the hard subprocess is identified with the physical 
$\psi$ state. In the CO model, the $c \bar{c}$ pair emerges from the hard process 
with quantum numbers different from those of the $\psi$ and evolves into the physical 
$\psi$ state by emitting one or more soft gluons. At LO, only the photon-gluon-fusion 
diagram, \mbox{$\gamma g \rightarrow \psi g$}, contributes to the direct photon cross 
section, as shown in Fig.~\ref{fig-fey}a).
Figure~\ref{fig-fey}b) shows the LO diagram for the resolved photon process
in the CS framework. A diagram for the direct photon process in the CO
framework is shown in Fig.~\ref{fig-fey}c).
Although a full next-to-leading-order (NLO) calculation of $\psi$ 
photoproduction is not available for all processes, the NLO corrections to the direct 
photon process, carried out in the framework of the CS model, have been 
calculated~\cite{pl:b348:657,*np:b459:3}.

The production of $\psi$ mesons has been measured in $p\bar{p}$  collisions by the 
CDF collaboration~\cite{prl:79:572,prl:79:578}. Predictions of the CS model, which 
for $p\bar{p}$ collisions exist only at LO in QCD, underestimate the data by factors 
of between 10 and 80. However, after adjustment of the corresponding matrix elements, 
this difference can be accounted for by the CO 
contributions~\cite{prl:76:4128,*pr:d54:4312,*pr:d60:119902}. Currently, the matrix 
elements governing the strength of this process cannot be calculated, 
but have to be determined from experiment. Since they are expected to be universal, the 
comparison of the values extracted from $\psi$ cross-section measurements in different 
environments constitutes a stringent test of this approach. 

The $J/\psi$ helicity distributions predicted by the CS and CO models have a different 
dependence on the $p_T$ of the $J/\psi$. Furthermore, the dependence of the $J/\psi$ 
polarisation on its transverse momentum is sensitive to the virtuality of the initial 
gluon in the photon~\cite{pl:b428:377}. Results from the CDF 
collaboration~\cite{prl:85:2886} show some discrepancies between the helicity 
measurements and predictions~\cite{pr:d57:4258} using  CO matrix elements extracted 
from the CDF cross-section data. 

The various $\psi$ photoproduction processes can be distinguished using the 
inelasticity variable, $z$, defined as:
\begin{equation}
z =  \frac{P \cdot p_{\psi} }{P \cdot q},
\label{eq-zdef}
\end{equation}
where $P$, $p_{\psi}$ and $q$ are the four-momenta of the incoming proton, the $\psi$ 
meson and the exchanged photon. In the proton rest frame, $z$ is the fraction of the 
photon energy carried by the $\psi$. Previous HERA 
data~\cite{zfp:c76:599,np:b472:3,epj:c25:25} have shown that the diffractive 
process populates the high-$z$ region, $z >$ 0.9. The direct and resolved photon 
processes are expected to dominate in the regions $0.2 \lesssim z < 0.9$ and 
$z \lesssim$ 0.2, respectively~\cite{hep-ph-0106120}.

In this study, $\psi$ mesons were identified using the decay mode 
$\psi \rightarrow \mu^+ \mu^-$ and were measured in the range 50 $< W
<$ 180 GeV, where $W$ is the $\gamma p$ centre-of-mass 
energy. The differential cross sections are given for 
$z >$ 0.1 and for different regions of transverse momentum, $p_T$, of the $J/\psi$. 
The $J/\psi$ helicity distributions in the ranges \mbox{$0.4 < z <$ 1} and 
$0.4 < z <$ 0.9 are presented and compared to model predictions with or without CO 
contributions.

\section{Theoretical models}
\label{sec-teomod}

\subsection{Leading--order Colour Singlet calculations}

The LO matrix element for the photon-gluon-fusion process, as computed in the CS 
framework, is singular for $z=1$ and $p_T = 0$~\cite{pr:d23:1521}. Therefore, the 
comparisons with these theoretical calculations are restricted to the region 
$p_T>1$~GeV.

Calculations of direct processes at LO in the CS model have been available for some 
time~\cite{pr:d23:1521}. In this paper, the data are compared to the LO prediction from 
Kr\"{a}mer et al.~\cite{pl:b348:657,*np:b459:3,hep-ph-0106120} (KZSZ (LO, CS)), including 
both direct and resolved processes. This calculation used the parton density functions 
(PDFs) GRV94 LO\cite{zfp:c67:433} for the proton and GRV LO\cite{pr:d45:3986,*pr:d46:1973} 
for the photon, the QCD scale parameter, $\Lambda^{(4)}_{\rm QCD}$, was set to 0.2~GeV and 
the factorisation and renormalisation scales were set to $\mu = 2 m_c$, where $m_c$, the 
charm-quark mass, was set to 1.5~GeV. Recently, the calculation has been extended to 
include predictions of the $J/\psi$ helicity-angle distributions\cite{pr:d57:4258} 
\mbox{(BKV (LO, CS))}.

In the CS framework the distributions of the $J/\psi$ helicity angle have been calculated 
by Baranov~\cite{pl:b428:377} for the direct photon process. This calculation uses the 
$k_T$-unintegrated gluon densities satisfying the BFKL\cite{jetp:45:199,*sovjnp:28:822} 
evolution equations. Compared to traditional (collinear) parton models, gluons 
have a transverse-momentum component (or virtuality), which results in an increase in the 
fraction of longitudinally polarised $J/\psi$ mesons as $p_T$ increases.

\subsection{Next--to--leading--order Colour Singlet calculation}

The NLO corrections to the direct photon process in the CS framework have been calculated 
by Kr\"{a}mer et al.~\cite{pl:b348:657,*np:b459:3} \mbox{(KZSZ (NLO, CS))}. This is the 
only NLO calculation currently available for any $J/\psi$ production process. The 
uncertainties in the cross sections arise from uncertainties in the non-perturbative QCD 
parameters. Upper bounds were obtained by setting \mbox{$m_c =$ 1.3 GeV} and the strong 
coupling constant, $\alpha_s(M_Z)$, to 0.121 in accordance with the 
MRST01~($\alpha_s\uparrow$)~\cite{epj:c23:73} set of proton PDFs. The lower bounds 
were obtained by setting \mbox{$m_c =$ 1.6 GeV} and \mbox{$\alpha_s(M_Z) =$ 0.117}, in 
accordance with the MRST01~($\alpha_s\downarrow$) set of PDFs. The dominant uncertainty is 
due to the variation of the charm-quark mass. For the calculation of the cross-section 
$d\sigma/dp_T^2$, the factorisation and renormalisation scales were set to the larger of 
$m_c/\sqrt{2}$ and $\left( \sqrt{m_c^2 + p_T^2} \right)/2$. For the prediction of the 
cross-sections $d\sigma/dz$, the factorisation and renormalisation scales were set to 
$m_c/\sqrt{2}$\cite{pl:b348:657,*np:b459:3}.

\subsection{Non--relativistic Quantum Chromodynamics calculations}

The LO calculation from Kr\"{a}mer et al. has also been extended to include the CO 
contributions, (KZSZ (LO, CS+CO)), from both direct and resolved photon 
processes~\cite{hep-ph-0106120}. The CO matrix elements were extracted by fitting the 
cross-section $d\sigma/dp_T^2$ for prompt $J/\psi$ production measured by 
CDF~\cite{prl:79:578}. The matrix elements for the hard subprocesses were computed at LO, 
while the CO matrix elements were corrected for initial- and final-state gluon radiation 
by a Monte Carlo (MC) technique~\cite{prl:76:4128,*pr:d54:4312,*pr:d60:119902}. The spread in the 
predictions is due to theoretical uncertainties in the extraction of the CO matrix elements 
obtained by comparing the values extracted by different groups; this spread is often larger 
than the error quoted by each individual group. This calculation has also been extended to 
predict the $J/\psi$ helicity-angle distributions\cite{pr:d57:4258} (BKV (LO, CS+CO)).

A LO calculation by Beneke, Schuler and Wolf~\cite{pr:d62:34004} (BSW (LO, CS+CO)) includes 
only the direct photon process for the CS and CO contributions. Here, the CO matrix elements 
were extracted from measurements by the CLEO collaboration~\cite{pr:d52:2661} on $B$ meson 
decays to $J/\psi$ mesons. The matrix elements extracted using the data from CLEO 
and CDF are consistent~\cite{hep-ph-0106120}. This calculation models the so-called shape 
functions that resum an infinite class of CO contributions that are important at high $z$. 
These functions are responsible for the decrease of the CO contributions towards $z=1$, due 
to the lack of phase space for gluon radiation. This treatment introduces an additional 
parameter into the model called the shape-function parameter which was varied in the range 
$300-500$~MeV, based on an evaluation\cite{pr:d62:34004} of the CLEO data.

Kniehl and Kramer~\cite{epj:c6:493} (KK (LO, CS+CO)), like Kr\"{a}mer et al., have 
calculated CS and CO terms in LO for both direct and resolved photon processes. The CO 
matrix elements were similarly extracted by fitting the $d\sigma/dp_T^2$ differential cross 
section for prompt $J/\psi$ production measured by CDF~\cite{prl:79:578}. The spread in the 
predictions is due to theoretical uncertainties in the extraction of the CO matrix elements. 
The calculation approximately takes into account dominant higher-order (HO) QCD effects and 
was performed in the $\overline{\mbox{MS}}$ renormalisation and factorisation scheme, using 
CTEQ4LO\cite{pr:d55:1280} and GRV LO as the proton and photon PDFs, respectively; the QCD 
scale parameter $\Lambda^{(4)}_{\rm QCD}$ was set to 296 MeV; common factorisation and 
renormalisation scales were used and were set to $\mu = \sqrt{4m_c^2+p^2_T}$ with 
$2 m_c = m_{J/\psi}$. 

\section{Experimental conditions}

The data were collected during the 1996 and 1997 running periods, when HERA operated with 
protons of energy $E_p~=~820~{\mbox{GeV}}$ and positrons of energy $E_e~=~27.5$~GeV, and 
correspond to an integrated luminosity of $38.0\pm0.6~\mbox{pb}^{-1}$. This represents more 
than a tenfold increase with respect to the previous ZEUS analysis~\cite{zfp:c76:599}. A 
detailed description of the ZEUS detector can be found 
elsewhere~\cite{zeus:1993:bluebook,pl:b293:465}. A brief outline of the components that are 
most relevant for this analysis is given below.

\Zctddesc\ZcoosysfnBeta

\Zcaldesc
~The timing resolution of the CAL is better than 1 ns for energy deposits greater
than 4.5 GeV.

The muon system consists of tracking detectors (forward, barrel and rear muon chambers: 
FMUON~\cite{zeus:1993:bluebook}, B/RMUON~\cite{nim:a333:342}), which are placed inside and 
outside a magnetized iron yoke surrounding the CAL and cover polar angles from 10$^\circ$ 
to 171$^\circ$.
The barrel and rear inner muon chambers cover polar angles from 34$^\circ$ to 135$^\circ$ and
from 135$^\circ$ to 171$^\circ$, respectively.  

The luminosity was determined from the rate of the bremsstrahlung process 
$e^{+} p \rightarrow e^{+} \gamma p$, where the photon was measured by a lead--scintillator 
calorimeter~\cite{desy-92-066,*acpp:b32:2025} located at $Z=-107$~m.

\section{Event selection and $\psi$ reconstruction}
\label{sec:selection}

The $\psi \rightarrow \mu^+ \mu^-$ candidates were selected using a three-level trigger 
system~\cite{zeus:1993:bluebook}. In the first-level trigger (FLT), the barrel and rear 
inner muon chambers, BMUI and RMUI, were used to tag the muons from $\psi$ decays by 
matching segments in the muon chambers with tracks in the CTD, as well as with energy 
deposits in the CAL consistent with the passage of a minimum ionising particle (m.i.p.). 
Events satisfying this regional matching and having tracks in the CTD pointing to the 
nominal interaction vertex were selected.

In the second-level trigger (SLT), the total energy in the calorimeter 
($E = \Sigma_i E_i$) and the $Z$ component of the momentum 
($p_Z = \Sigma_i E_i \cos \theta_i$) were calculated. The sums run over all calorimeter 
cells $i$ with an energy, $E_i$, and polar angle, $\theta_i$, measured with respect to 
the nominal vertex. To remove proton-gas interactions, events with the ratio $p_Z/E$ 
greater than 0.96 were rejected. The cosmic-ray background was partially rejected at the 
SLT by using the time differences of energy deposits in the upper and the lower halves 
of the BCAL.

In the third-level trigger (TLT), a muon candidate was selected when a track found in 
the CTD matched both a m.i.p. in the CAL and a track in the inner muon chambers. An 
event containing a muon candidate in the rear (barrel) region was accepted if the 
momentum (transverse momentum) of the CTD track exceeded 1 GeV.

In the offline analysis, the TLT algorithm was again applied to the results of the full 
event reconstruction. In addition, the tracks corresponding to the two muons from the 
$\psi$ decay had to satisfy several criteria. One track was matched to both a m.i.p.\ 
cluster in the CAL and a track in the inner muon chambers. This track was required to 
have a momentum greater than 1.8 GeV if it was in the rear region, or a transverse 
momentum greater than 1.4 GeV if in the barrel region. The other muon track was matched 
to a m.i.p.\ cluster in the CAL and was required to have a transverse momentum of greater 
than 0.9 GeV. Both tracks were restricted to the pseudorapidity region 
\mbox{$|\eta| < 1.75$}. To reject cosmic rays, events in which the angle between the 
two muon tracks was larger than $174^{\circ}$ were removed. Events were also required to 
have an energy deposit greater than 1~GeV in a cone of $35^{\circ}$ around the forward 
direction (excluding calorimeter deposits due to the decay muons). Elastically produced 
$\psi$ mesons were thus excluded.

The kinematic region considered was defined by the inelasticity variable $z$, 
given in Eq.~(\ref{eq-zdef}), and by the photon-proton centre-of-mass energy: 
\begin{equation}
W^2 = (P+q)^2; \nonumber
\end{equation}
$W$ and $z$ were computed from 
\begin{equation}
W^2 = 2 E_p (E-p_Z) \nonumber
\end{equation}
and
\begin{equation}
z = \frac{(E-p_Z)_{\psi}}{E-p_Z}, \nonumber
\end{equation}
where $E-p_Z = (E-p_Z)_{\rm had} + (E-p_Z)_{\psi}$. The quantity $(E-p_Z)_{\rm had}$ 
is the sum over the hadronic final state, calculated using all CAL cells excluding 
those belonging to the decay-muon clusters; $(E-p_Z)_{\psi}$ was calculated using 
the $\psi$ decay tracks measured by the CTD.

The events were required to have $E-p_Z < 20$~GeV, which restricts $W$ to be less than 
\mbox{180 GeV} and $Q^2 \lesssim 1$~GeV$^2$, with a median value of $\sim 10^{-4}$~GeV$^2$. 
The elimination of deep inelastic scattering events was independently confirmed by 
searching for scattered positrons in the CAL\cite{nim:a365:508}; none was found. As the 
analysis uses only the B/RMUON, the polar angle coverage of these detectors restricts $W$ 
to be greater than 50 GeV.

\section{Monte Carlo models}
\label{sec:mc}

The production of $\psi$ mesons from direct interactions was simulated using the 
HERWIG~5.8~\cite{cpc:67:465} MC generator, which generates events according 
to the LO diagrams of the photon-gluon-fusion process, $\gamma g \rightarrow \psi g$, as 
computed in the framework of the CS model. The hadronisation process is simulated by the 
cluster model~\cite{np:b238:492}. Events were generated in the range of $Q^2$ starting 
from the kinematic limit ($\approx 10^{-10}$ GeV$^2$) up to 10~GeV$^2$. Events were 
generated for $z<0.95$ to avoid a singular phase-space region. The GRV94 LO PDF for the 
proton was used. The HERWIG MC sample was reweighted in $p_T$ and $W$ to the data.

The production of $\psi$ mesons from resolved photon interactions was simulated using the 
PYTHIA 6.146~\cite{cpc:43:367,*cpc:46:43} MC generator (resolved photon interactions for 
$\psi$ production are not implemented in HERWIG). The GRV94 LO and GRV LO PDFs were used 
for the proton and photon, respectively. The matrix elements for the resolved photon 
processes were computed at LO in the framework of the CS model.

Diffractive production of $\psi$ mesons with proton dissociation was simulated with the 
EPSOFT~\cite{thesis:kasprzak:1994} MC generator, which has been tuned to describe such 
processes at HERA~\cite{thesis:adamczyk:1999}. Data in the region $0.9<z<1$ were used to 
determine the dependence of the cross section on the invariant mass of the dissociative 
system, on the photon-proton centre-of-mass energy and on the $p_T^2$ of the $J/\psi$ 
meson. 

\section{Signal determination and cross-section calculation}
\label{sec:sigxsec}

\begin{figure}
\epsfig{file=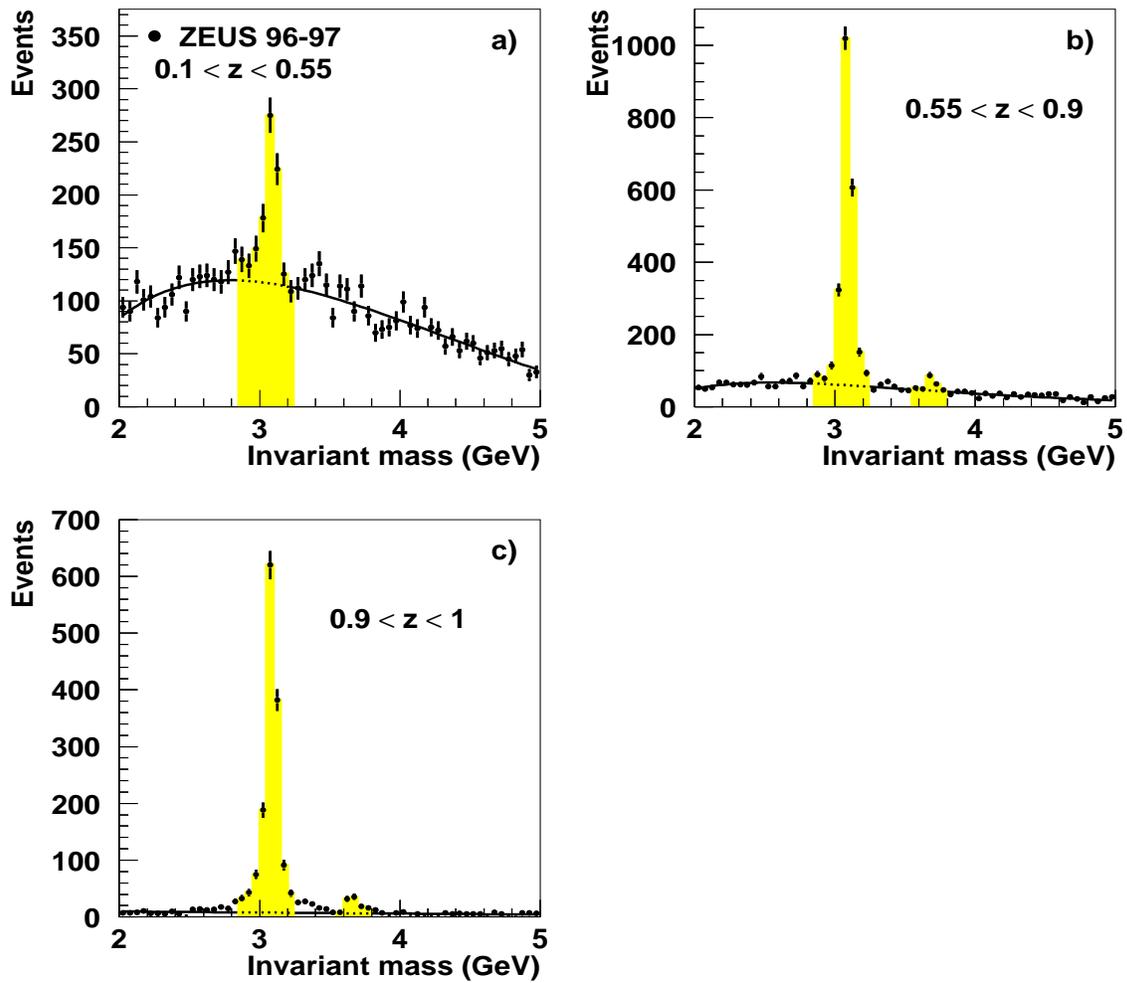,
height=15cm,width=15cm,clip=,angle=0,silent=}
\vspace{-0.5cm}
\caption{The invariant-mass spectrum measured in the 
region \mbox{50 $< W <$~180~GeV} for a) 0.1 $< z <$ 0.55, b) 0.55 $< z <$ 0.9 
and c) 0.9 $< z <$ 1. The signal regions are shown as the shaded bands and 
the background as the continuous line.}
\label{fig-mmumu}
\end{figure}

The invariant-mass spectra of the muon pairs measured for 50 $< W <$ 180 GeV and three 
$z$ ranges are shown in \fig{mmumu}. The $J/\psi$ is clearly seen in all $z$ ranges. The 
higher $z$ ranges (Figs.~\ref{fig-mmumu}b) and \ref{fig-mmumu}c)) also show a 
$\psi^\prime$ peak.

To estimate the number of events in the signal regions (the mass ranges 2.9 to 3.25~GeV 
and 3.6 to 3.8~GeV for the $J/\psi$ and $\psi^\prime$, respectively), an accurate 
description of the combinatorial background was necessary. This was estimated by fitting 
the invariant-mass distribution of the data outside the corresponding windows of the 
$\psi$ mesons, using a function which is the product of a second-order polynomial and an 
exponential. 

The data were corrected bin by bin for geometric acceptance, detector, trigger and 
reconstruction efficiencies, as well as for detector resolution. The correction 
factor, as a function of an observable ${\mathcal O}$ in a given 
bin $i$, is $C_i({\mathcal O})=N_i^{\rm gen}({\mathcal O})/N_i^{\rm rec}({\mathcal O})$.
The variable $N_i^{\rm gen}({\mathcal O})$ is the number of generated events and 
$N_i^{\rm rec}({\mathcal O})$ is the number of reconstructed 
events passing the selection requirements detailed in Section~\ref{sec:selection}. 
Both numbers were computed using the MC generators described in Section~\ref{sec:mc}. 
In this $W$ range and the three regions of $z$, 0.1 $< z <$ 0.4, 0.4 $< z <$ 0.9 and 
0.9 $< z <$1, the acceptance (defined as $1/C_i({\mathcal O})$) was typically 30$\%$  
and always above 10$\%$.

\begin{figure}
\epsfig{file=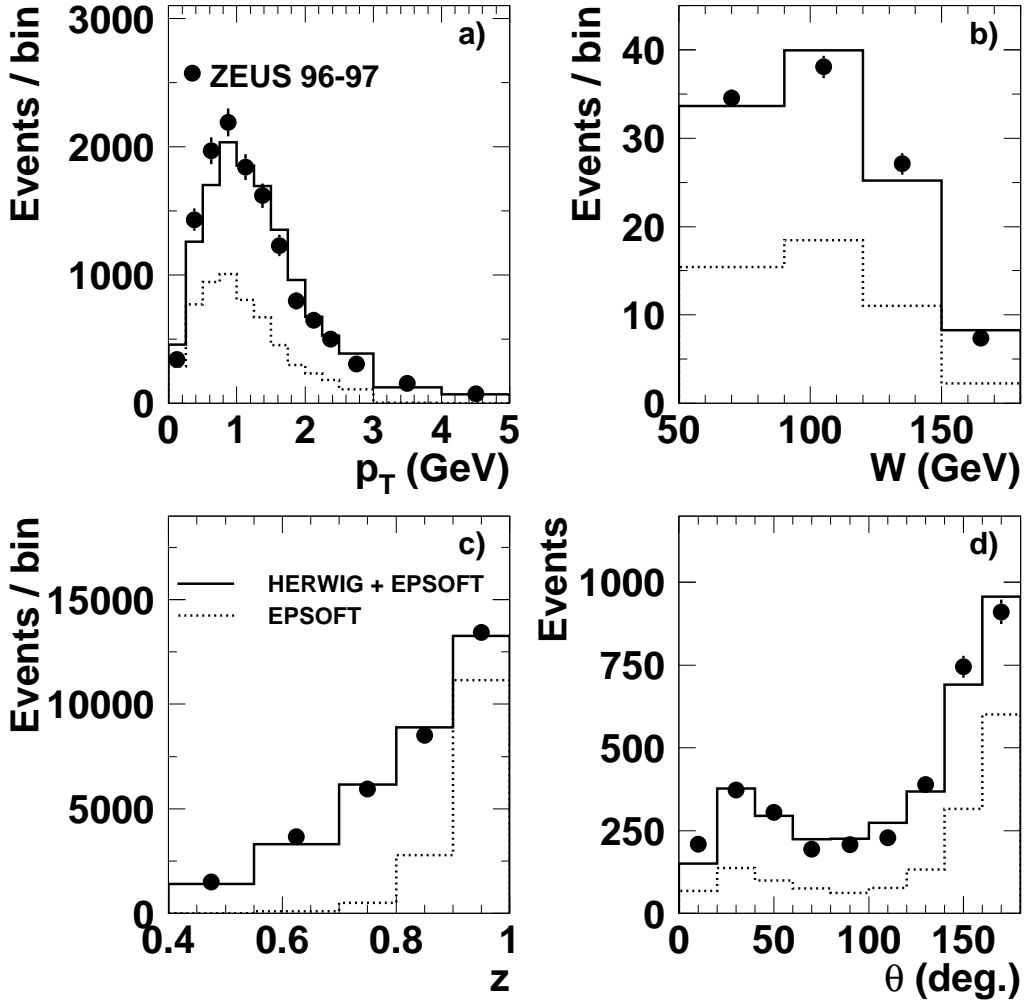,
height=15cm,width=15cm,clip=,angle=0,silent=}
\vspace{-0.5cm}
\caption{Number of events reconstructed in the kinematic region $z >$ 0.4 and 
50 $< W <$ 180 GeV plotted against a) $J/\psi$ $p_T$, b) $W$ c) inelasticity, $z$ and 
d) $J/\psi$ polar angle, $\theta$. The data distributions are shown as the points with 
statistical errors only. The simulated EPSOFT diffractive proton-dissociation background is 
shown as the dotted lines. The solid lines show the prediction of the sum of the HERWIG and 
EPSOFT generators. The combined MC has been area normalised to the data.
The HERWIG MC sample was reweighted in $p_T$ and $W$ to the data.}
\label{fig-gloevtpro}
\end{figure}

For \mbox{0.9 $< z <$ 1}, the events are largely diffractive. Therefore, the analysis of 
inelastic production was restricted to the region $0.1 < z < 0.9$. The remaining 
contamination from diffractive processes was estimated by fitting the relative fractions 
of HERWIG and EPSOFT MC event samples to the data. A $\chi^2$ fit was performed to the 
inelasticity distribution in the region \mbox{0.4 $< z <$ 1} and three $p_T$ ranges: 
\mbox{$0 < p_T < 1$}, $1< p_T < 2$ and \mbox{$p_T > 2$ GeV}. Figure~\ref{fig-gloevtpro} shows, 
for events in the region \mbox{50 $< W<$ 180 GeV} and \mbox{0.4 $< z <$ 1}, that the 
resulting mixture of $56\%$ HERWIG and $44\%$ EPSOFT gives a reasonable description of 
the relevant $J/\psi$ event observables. For \mbox{$0.4 < z < 0.9$}, with no $p_T$ cut, 
the diffractive contribution, as estimated with the EPSOFT MC, was 17\%, concentrated at 
low $p_T$. The diffractive contribution was subtracted bin by bin for all cross-section 
measurements.

Resolved photon processes are also present in the region of the cross-section measurement. 
In the region $0.1 < z < 0.9$, the resolved photon component was estimated by fitting the 
relative fractions of direct and resolved photon events in the MC samples to the 
inelasticity distribution in the data. The fraction of resolved photon events is $5\%$, 
reaching up to $50\%$ for \mbox{$0.1 < z < 0.4$}, in good agreement with theoretical 
expectations~\cite{hep-ph-0106120}. In this low $z$ range the contribution from $B$ decays 
can be as large as $25\%$; it is negligible at higher $z$. The photoproduction cross 
section was obtained from the measured electron-proton cross section by dividing by the 
integrated effective photon flux~\cite{zfp:c76:599}.

\section{Systematic uncertainties}
\label{sec:syst}

A detailed study of possible sources of systematic uncertainties was
carried out for all measured differential cross sections. The following 
sources were considered:

\begin{itemize}

\item muon-chamber efficiencies: the BMUI and RMUI muon-chamber efficiencies were extracted 
      from the data using muon pairs coming from elastic $J/\psi$ events and from the 
      process $\gamma \gamma \rightarrow \mu^+ \mu^-$. For $p_T>1.4$~GeV, the product of 
      the geometrical and chamber efficiency for the BMUI is greater than $30\%$, reaching 
      $60\%$ at high transverse momentum. For $p>1.8$~GeV, this product for the RMUI is greater than 
      $45\%$, reaching $70\%$ at high momentum. The associated uncertainty of about $\pm 7\%$ is 
      independent of the phase-space region;  

\item analysis cuts: this class comprises the systematic uncertainties due to the 
      uncertainties in the measurement of momentum, transverse momentum and pseudorapidity 
      of the muon decay tracks. Each cut was varied within a range determined by the 
      resolution in the appropriate variable. The pseudorapidity contribution gave an 
      uncertainty below $\pm 1\%$, while variation of the appropriate momentum or 
      transverse momentum cut gave a $\pm 1.5\%$ contribution;

\item CAL energy scale: CAL energy measurements were used in the $W$ and $z$ reconstruction. 
      This leads to a systematic uncertainty in the measured cross section due to the 
      $\pm 3\%$ uncertainty on the energy scale of the CAL. This effect was investigated by 
      varying the quantity $(E-p_Z)_{\rm had}$ by $\pm 3\%$ in the MC sample, leading to a 
      variation of the cross sections of $\pm 6\%$ for \mbox{$0.1 < z < 0.4$}. Integrated 
      over $z$, the effect is typically below $\pm 2\%$;

\item CAL energy resolution: the $W$ and $z$ resolutions are dominated by the CAL energy 
      resolution through the quantity $(E-p_Z)_{\rm had}$. The $(E-p_Z)_{\rm had}$ 
      resolution in the MC was smeared event by event by $\pm 20\%$. This estimated any 
      possible mismatch between the $(E-p_Z)_{\rm had}$ resolution in the data and MC 
      simulation, giving an uncertainty of $\pm 3\%$;

\item diffractive subtraction: the fraction of HERWIG and EPSOFT MC events, fixed by the 
      fitting procedure described in Section~\ref{sec:sigxsec}, is known to a precision 
      limited by the number of $J/\psi$ events in the data and the process modelling. All 
      possible fractions giving a $\chi^2$ in the interval $[\chi^2_{min},\chi^2_{min}+1]$ 
      were considered and the largest change in the cross section was quoted as the 
      systematic uncertainty. This gave an uncertainty which was at most $\pm 2\%$ at high 
      $z$ and low $p_T$, where the diffractive contribution peaks;

\item diffractive simulation: the EPSOFT MC simulation parameters were varied within ranges 
      allowed by the comparison between the data and the EPSOFT MC simulation in the region 
      $0.9<z<1$. The fraction of HERWIG and EPSOFT MC events was re-evaluated. This gave an 
      uncertainty which was at most $\pm 2.5\%$ at high $z$ and low $p_T$;

\item $p_T$ and $W$ spectra: the $p_T$ and $W$ spectra of the $J/\psi$ meson in the HERWIG MC 
      simulation were varied within ranges allowed by the comparison between the data and the 
      simulation and the correction factors re-evaluated. This gave an uncertainty of $\pm 2\%$;

\item helicity parameterisation uncertainty: in the HERWIG MC, the helicity parameter 
      $\alpha$ is set to 0. As a systematic check, the helicity in the HERWIG MC was 
      reweighted according to the upper and lower limits of error in the measured 
      distribution and the correction factors re-evaluated. This gave an uncertainty of 
      $\pm 5\%$, independent of the phase-space region.

\end{itemize}

All of the above individual sources of systematic uncertainty were added in quadrature. The 
following sources resulted in an overall shift of the cross section and were therefore 
treated separately:

\begin{itemize}

\item the integrated luminosity determination gave an uncertainty of $\pm 1.6\%$;

\item the branching ratio of $J/\psi \rightarrow \mu^+\mu^-$ gave an uncertainty of 
      $\pm 1.7 \%$~\cite{pr:d66:10001}.

\end{itemize}

\section{Results}

\subsection{Total cross-section measurement at high $z$}
\label{sec:highz}

A cross-section measurement in the region $z>0.9$ is particularly interesting because the 
CO mechanism is expected to contribute significantly at high $z$~\cite{hep-ph-0106120}, 
whereas for \mbox{$z<0.9$} the sensitivity to this production mechanism is reduced. In 
particular, at large $z$, the impact of the soft-gluon emission on the hadronisation of 
CO $c\bar{c}$ pairs is not well understood~\cite{hep-ph-0106120,pr:d62:34004,pl:b408:373}. 
In the region $0.9 < z < 1$, the $z$ resolution is comparable to the width of the $z$ 
interval and the diffractive process is dominant; hence the separation of the direct and 
diffractive components is not reliable. Therefore, only the visible cross section in the 
region \mbox{$0.9 < z < 1$}, including the diffractive component, is given. Furthermore, 
due to the requirement of an energy deposit in the direction of the outgoing proton, 
necessary to remove the elastic component, only diffractive events with a high-enough 
invariant mass, $M_N$, of the final-state hadronic system were included here. Monte Carlo 
studies showed that the requirement of an energy deposit exceeding 1 GeV in a 35$^\circ$ 
cone around the outgoing proton direction corresponds to a threshold in $M_N$ of 4.4 GeV, 
above which all correction factors were independent of $M_N$. The following cross sections,
corresponding to the phase--space region defined by 50 $< W <$ 180 GeV, 0.9 $< z <$ 1,
$M_N >$ 4.4 GeV were obtained:
\begin{eqnarray}
\sigma_{J/\psi}(p_T>0~\mbox{GeV}) &=&
45.7~\pm~1.3~\mbox{(stat.)}~^{+9.4}_{-4.6}~\mbox{(syst.) nb};
\nonumber \\
\sigma_{J/\psi}(p_T>1~\mbox{GeV}) &=&
24.5~\pm~0.9~\mbox{(stat.)}~^{+4.3}_{-2.5}~\mbox{(syst.) nb};
\nonumber \\
\sigma_{J/\psi}(p_T>2~\mbox{GeV}) &=&
6.5~\pm~0.5~\mbox{(stat.)}~^{+0.8}_{-0.7}~\mbox{(syst.) nb}. 
\nonumber
\end{eqnarray}
The uncertainties on the value of the $M_N$ threshold and the CAL energy resolutions 
are of similar importance and dominate the systematic uncertainty.

\subsection{Measurement of $\psi^{\prime}$ production}
\label{sec-pdifbkg}

The production of $\psi^{\prime}$ with subsequent decay to $J/\psi$ has been measured using 
the rates of $\psi^{\prime} \rightarrow \mu^+\mu^-$ and $J/\psi \rightarrow \mu^+\mu^-$. The 
$\psi^{\prime}$ to $J/\psi$ cross-section ratio was determined in the region 
\mbox{0.55 $< z <$ 0.9} with no $p_T$ cut on the $\psi$ mesons. The range 
\mbox{$0.4 < z < 0.55$} was not included, because it has a large non-resonant background. 
The $\psi^{\prime}$ to $J/\psi$ cross-section ratio was computed in bins of $p_T$, $W$ and $z$ 
from
\begin{eqnarray}
\frac{\sigma_i(\psi^\prime)}{\sigma_i(J/\psi)} = 
\frac{N_i^{2S}}{N_i^{1S}} \cdot
\frac{C_i^{2S}}{C_i^{1S}} \cdot
\frac{Br^{\mu}}{Br^{\mu^{\prime}}} \cdot 
\left(1-\frac{N_i^{2S}}{N_i^{1S}} ~
\frac{C_i^{2S}}{C_i^{1S}} ~
\frac{Br^{\mu}}{Br^{\mu^{\prime}}} ~
Br^{\prime} \right)^{-1} ,
\label{eq-sigrat}
\nonumber
\end{eqnarray}
where, for the considered bin $i$, $N^{1S}_i$ ($N^{2S}_i$) is the number of $J/\psi$ 
($\psi^{\prime}$) events observed, $C^{1S}_i$ ($C^{2S}_i$) is the correction factor 
(see Section~\ref{sec:sigxsec})  computed using HERWIG MC $J/\psi$ ($\psi^{\prime}$) events, 
$Br^{\mu}$ ($Br^{\mu^{\prime}}$) is the $J/\psi$ ($\psi^{\prime}$) muonic 
branching ratio and $Br^{\prime}$ is the $\psi^{\prime} \rightarrow J/\psi~X$ 
branching ratio. The values used were $Br^{\mu} = (5.88\pm0.10)\%$, 
$Br^{\mu^{\prime}} = (0.70\pm0.09)\%$ and $Br^{\prime} = (55.7\pm2.6)\%$~\cite{pr:d66:10001}.

\begin{figure}
\epsfig{file=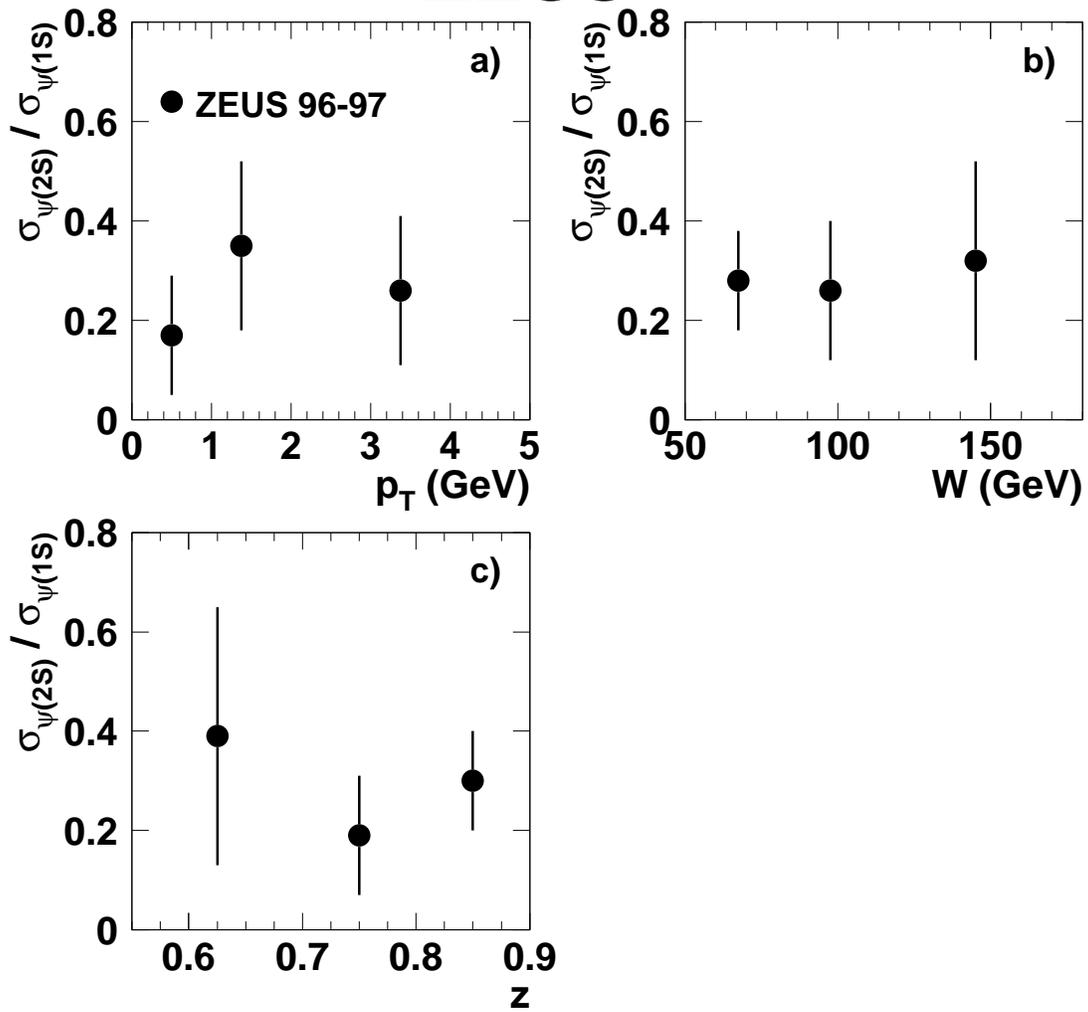,
height=15cm,width=15cm,clip=,angle=0,silent=}
\vspace{-0.5cm}
\caption{$\psi^{\prime}$ to $J/\psi$ cross-section ratios as a function of a)
$p_T$, b) $W$, and c) $z$, measured for $50 < W < 180$ GeV and $0.55 < z < 0.9$. 
The uncertainties are statistical only.}
\label{fig-2sto1sevtr}
\end{figure}

With this technique, the cross-section ratio is corrected for the 
$\psi^{\prime} \rightarrow J/\psi~(\rightarrow \mu^+\mu^-)~X$ cascade decay. The results are 
shown in Fig.~\ref{fig-2sto1sevtr} and listed in Table~\ref{Tab:psipratio}; 
all cross-section ratios are consistent with being 
independent of the kinematic variable, as expected if the underlying production mechanisms 
for the $J/\psi$ and $\psi^{\prime}$ are the same. For the range $0.55 < z < 0.9$ and 
\mbox{$50 < W < 180$ GeV}, the $\psi^{\prime}$ to $J/\psi$ cross-section ratio is
\begin{equation}
\frac{\sigma(\psi^{\prime})}{\sigma(J/\psi)} =
0.33 \pm 0.10~(\mbox{stat.}) ^{+0.01}_{-0.02}~(\mbox{syst.}),
\end{equation}
in agreement with the expectation of the LO CS model of 0.24~\cite{pl:b348:657,*np:b459:3}. 

Even though the NLO corrections to the CS model for $J/\psi$ production are known to be 
large, similar large NLO corrections are expected to affect the rate of $\psi^{\prime}$ 
production~\cite{priv:kraemer:2002}. Hence the $\psi^{\prime}$ to $J/\psi$ 
cross-section ratio at NLO is not expected to differ significantly from that at LO. From 
the cross-section ratio and the $\psi^{\prime} \rightarrow J/\psi X$ branching 
ratio~\cite{pr:d66:10001}, it is estimated that the observed cross section for $J/\psi$ mesons 
is increased by \mbox{(18.4$\pm$5.6~(stat.))$\%$} due to $J/\psi$ mesons originating from 
$\psi^{\prime}$ cascade decays. This is consistent with the expected value of 
$15\%$\cite{pl:b348:657,*np:b459:3}, which has been added to all predictions of the 
$J/\psi$ differential cross sections presented in this paper.

\subsection{Measurement of inelastic $J/\psi$ cross sections}

\begin{figure}
\epsfig{file=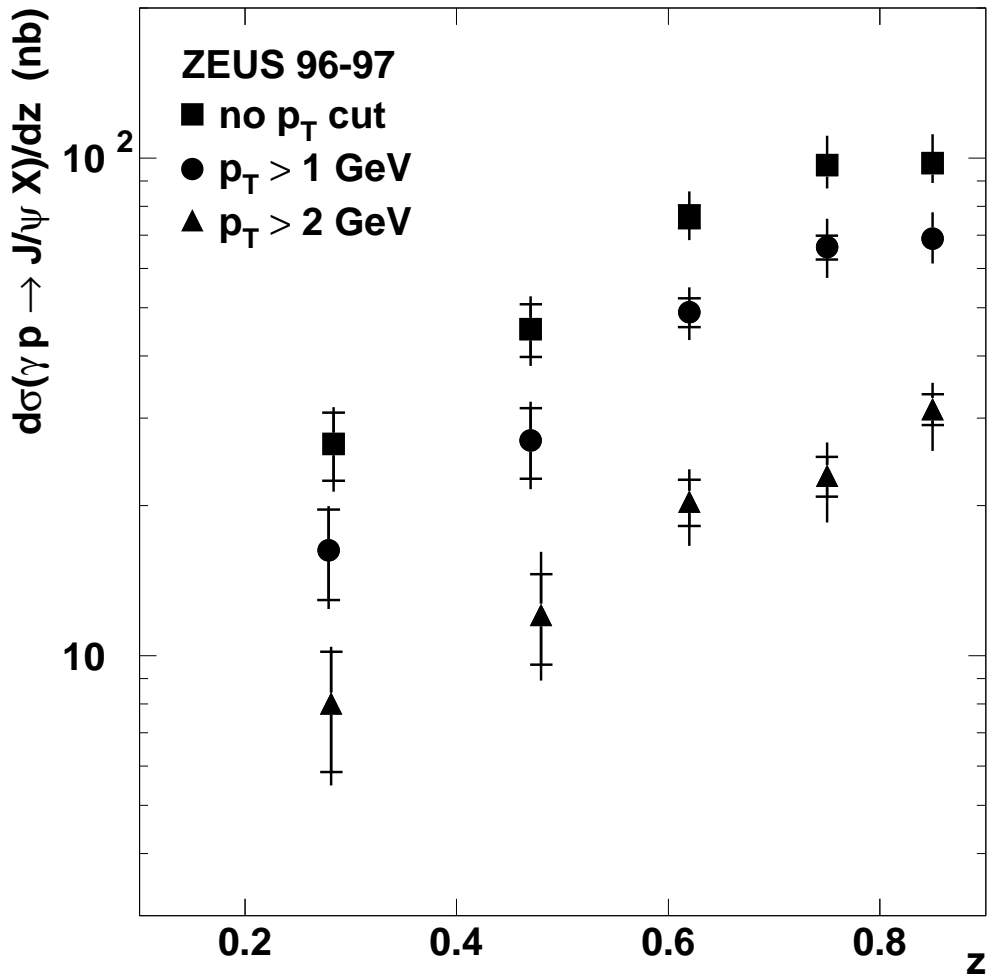,
height=15cm,width=15cm,clip=,angle=0,silent=}
\vspace{-0.5cm}
\caption{Differential cross-sections $d\sigma/dz$ for $50 < W < 180$ GeV and different $p_T$ 
selections: no $p_T$ cut (squares), $p_T > 1$ (circles), and $p_T > 2$ GeV (triangles). The 
inner error bars show the statistical uncertainty; the outer bars show the statistical and 
systematic uncertainties added in quadrature.}
\label{fig-dsdzmultipt}
\end{figure}

\begin{figure}
\epsfig{file=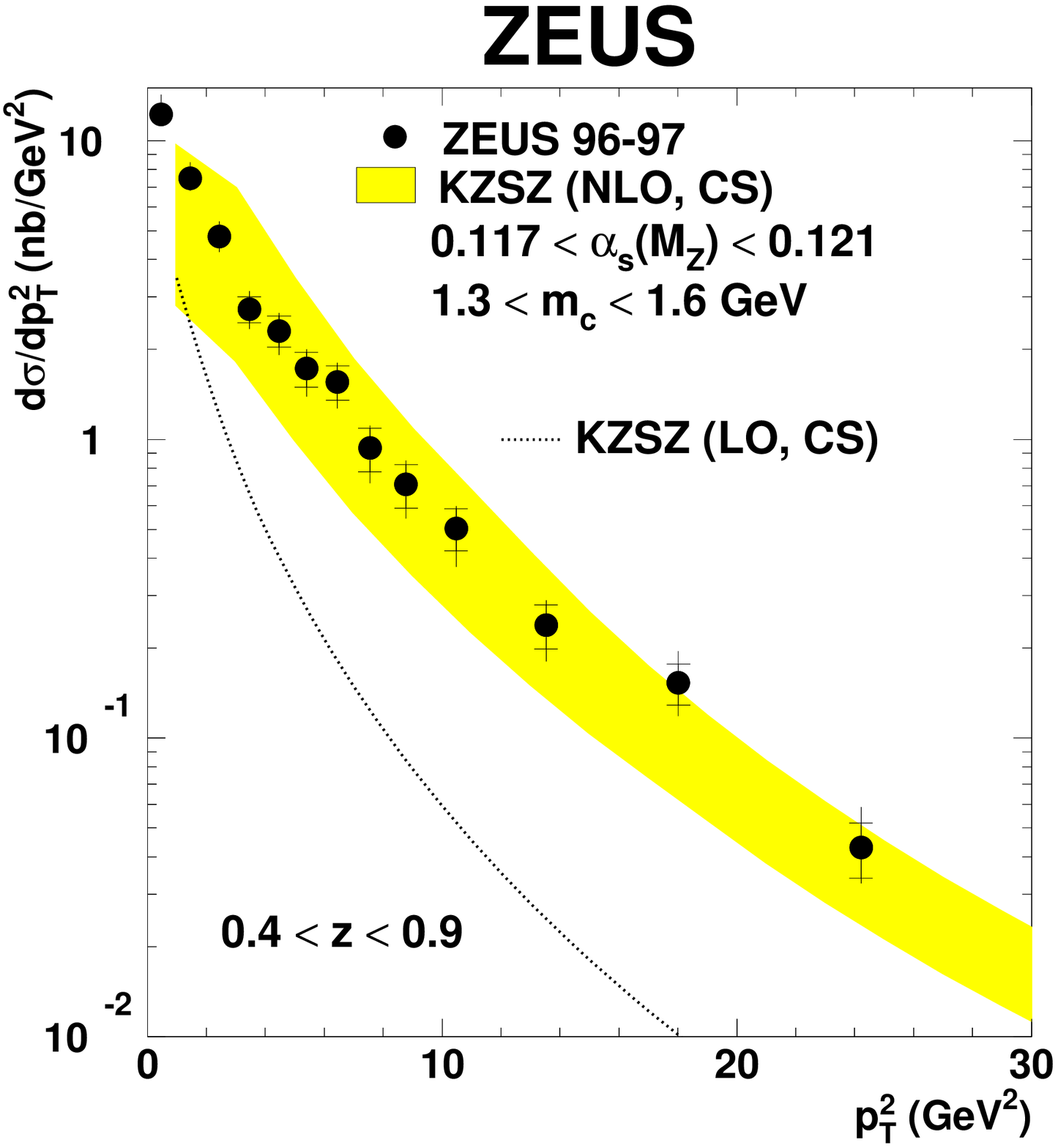,
height=15cm,width=15cm,clip=,angle=0,silent=}
\vspace{-0.5cm}
\caption{$J/\psi$ differential cross-section $d\sigma/dp_T^2$ for $50 < W < 180$ GeV and 
$0.4 < z < 0.9$. The inner error bars show the statistical uncertainty; the outer bars show 
the statistical and systematic uncertainties added in quadrature. The data points are 
compared to the prediction \mbox{KZSZ (NLO, CS)} (shaded band) including only the direct 
photon process. 
The spread in the prediction is due to uncertainties on the charm-quark mass and 
on the QCD scale parameter, $\Lambda_{QCD}$. The dotted line represents the LO prediction 
\mbox{KZSZ (LO, CS)}.
A 15 \% contribution has been added to the predictions to account for $J/\psi$ mesons 
originating from $\psi^{\prime}$ cascade decays.}
\label{fig-dsdpt2}
\end{figure}

\begin{figure}
\epsfig{file=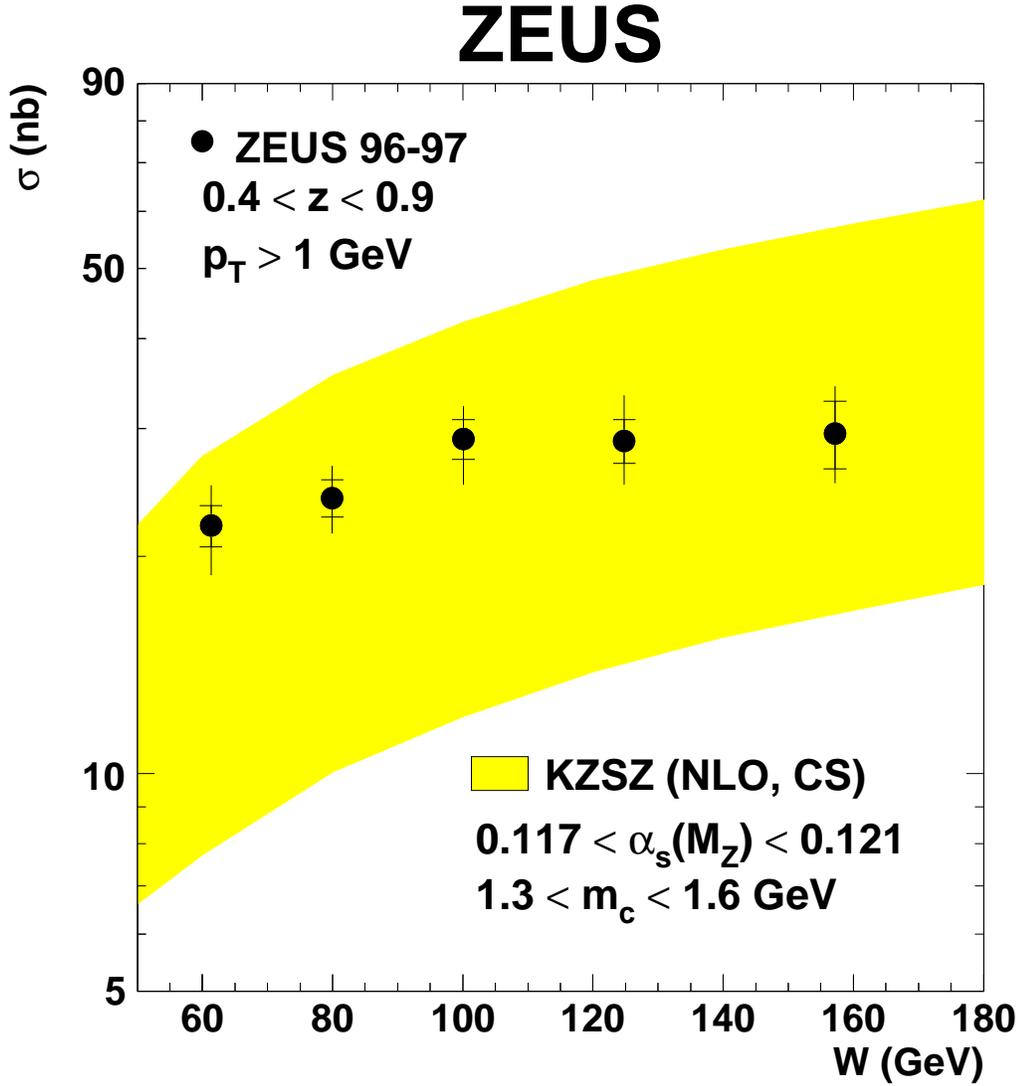,
height=15cm,width=15cm,clip=,angle=0,silent=}
\vspace{-0.5cm}
\caption{Cross section as a function of $W$ for $p_T > 1$~GeV and \mbox{0.4 $<z<$ 0.9}. 
The inner error bars show the statistical uncertainty; the outer bars the statistical and 
systematic uncertainties added in quadrature. The shaded band shows the prediction 
\mbox{KZSZ (NLO, CS)}. The spread in the prediction is due to 
uncertainties on the charm-quark mass and on the QCD scale parameter, $\Lambda_{QCD}$.
A 15 \% contribution has been added to the prediction to account for $J/\psi$ mesons 
originating from $\psi^{\prime}$ cascade decays.}
\label{fig-dsdw}
\end{figure}

\begin{figure}
\epsfig{file=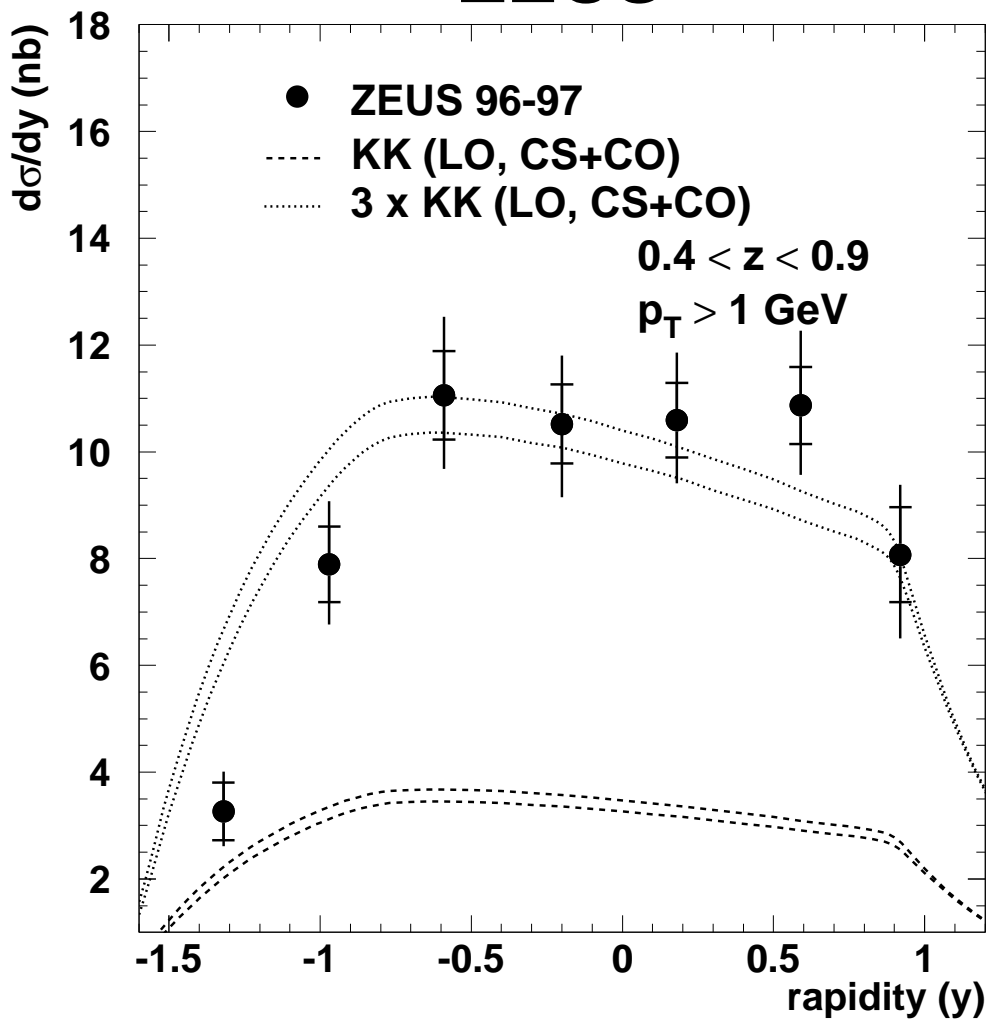,
height=15cm,width=15cm,clip=,angle=0,silent=}
\vspace{-0.5cm}
\caption{$J/\psi$ differential cross-section $d\sigma/dy$ for $50 < W < 180$ GeV, 
\mbox{$0.4 < z < 0.9$} and $p_T > 1$ GeV. The inner error bars show the statistical uncertainty; 
the outer bars show the statistical and systematic uncertainties added in quadrature. The 
range of the \mbox{KK (LO, CS+CO)} prediction is shown as the dashed
lines. The dotted lines 
show the same prediction scaled up by a factor of three. The spread in the predictions is 
due to theoretical uncertainties in the extraction of the CO matrix elements.
A 15 \% contribution has been added to the predictions to account for $J/\psi$ mesons 
originating from $\psi^{\prime}$ cascade decays.}
\label{fig-dsdy}
\end{figure}

The differential cross-sections $d\sigma/dz$ for three different regions in $p_T$ and 
\mbox{0.1 $< z <$ 0.9} are shown in 
\mbox{Figs.~\ref{fig-dsdzmultipt},\ref{fig-dsdzpt1} and \ref{fig-dsdzpt2}} 
and listed in \mbox{Tables \ref{Tab:dsdzpt0},\ref{Tab:dsdzpt1} and 
\ref{Tab:dsdzpt2}}. All data sets 
show a cross section increasing with $z$. The differential cross-section $d\sigma/dp_T^2$,
measured in the region \mbox{0.4 $< z <$ 0.9}, is shown in Fig.~\ref{fig-dsdpt2} and 
listed in Table~\ref{Tab:dsdpt2}. The measurement 
extends to $p_T^2 \sim 24~{\rm GeV^2}$, where the cross section has fallen by two orders of 
magnitude. The cross-section has a function of $W$, for 
\mbox{$p_T>1$~GeV} and 0.4 $< z <$ 0.9, is given in Table~\ref{Tab:sversusW} 
and shown in Fig.~\ref{fig-dsdw}. 
The differential cross-section $d\sigma/dy$ in the region $0.4<z<0.9$ and $p_T>1$ GeV is 
shown in \fig{dsdy} and listed in Table~\ref{Tab:dsdy}. 
The rapidity, $y$, of the $J/\psi$ is given by
\[ y \equiv \frac{1}{2} \ln \frac{E+p_Z}{E-p_Z}, \]
where $E$ and $p_Z$ are the energy and $Z$ component of the momentum of the $J/\psi$, 
respectively.

The helicity distribution can be parameterised as
\begin{equation}
\frac{dN}{d \cos \theta^*} \propto 1 + \alpha \cos^2 \theta^* .
\label{eq-helieq}
\end{equation}
The helicity-parameter $\alpha$ was determined by reweighting the HERWIG MC 
$dN/d\cos \theta^*$ generator-level distribution according to Eq.~\eq{helieq} for different 
values of $\alpha$. The $\chi^2$ for the $dN/d|\cos \theta^*|$ distribution in data and MC 
was then calculated for each value of $\alpha$ in the MC and the minimum $\chi^2$ gave the 
central value of $\alpha$. The procedure was repeated for each $p_T$ bin in the range 
$1 < p_T < 5$ GeV. The systematic uncertainties were negligible with respect to the error 
obtained from the $\chi^2$ fit. 

\begin{figure}
\unitlength1cm \begin{picture}(18.,15.)
\includegraphics{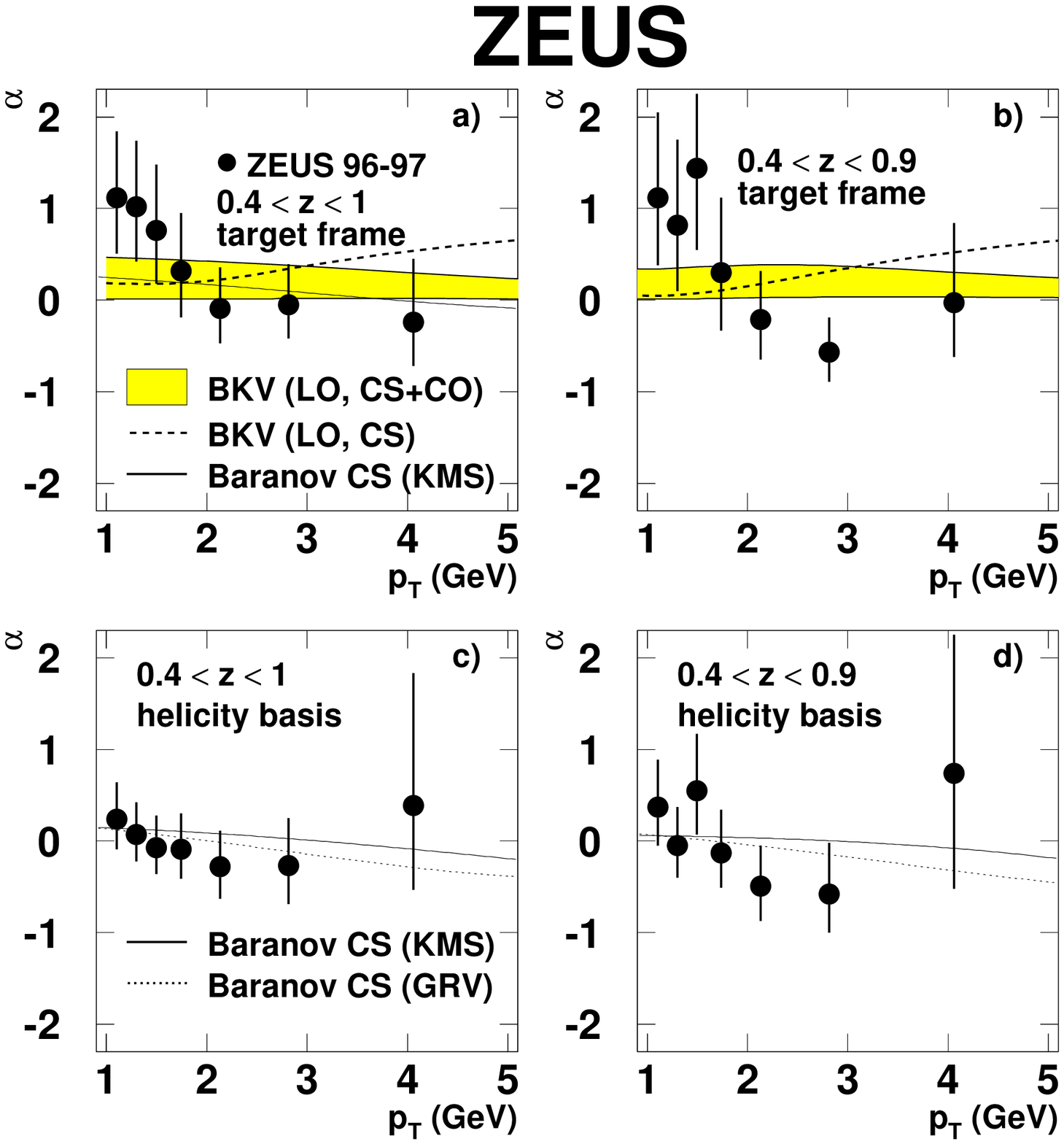}
\end{picture}
\vspace{-0.5cm}
\caption{$J/\psi$ helicity parameter, $\alpha$, as a function of $p_T$ for 
\mbox{$50 < W < 180$ GeV} a) and c) $0.4 < z < 1$; b) and d) $0.4 < z <0.9$. 
The error bars correspond to the total experimental uncertainties.
The results for the target frame are shown in a) and 
b) and the results for the helicity basis frame are shown in c) and d). In a) and b), the 
prediction of BKV (LO CS+CO) is shown as the shaded band, while the prediction from the 
BKV (LO, CS) model is shown as the dashed line. In a), c) and d), the data are compared with 
the predictions of Baranov using the GRV (dotted line) and KMS (solid line) unintegrated 
parton densities.}
\label{fig-helitarg}
\end{figure}

Figure~\ref{fig-helitarg} shows the measured parameter $\alpha$ plotted as a function of the 
$p_T$ of the $J/\psi$. In Fig.~\ref{fig-helitarg}a) and b), the quantisation axis is chosen 
to be the opposite of the incoming proton direction in the $J/\psi$ rest frame, $\theta^*$ is 
the opening angle between the quantisation axis and the $\mu^+$ direction of flight in the 
$J/\psi$ rest frame and $\alpha$ is the helicity parameter. This frame is known as the 
``target frame''. The parameter $\alpha$ was determined in bins of $p_T$, for 
\mbox{$p_T >$ 1 GeV} and 0.4 $< z <$ 1 (Fig.~\ref{fig-helitarg}a)). The parameter $\alpha$ 
was also measured in the range $0.4 < z < 0.9$ (Fig.~\ref{fig-helitarg}b)), where the 
diffractive contamination is reduced. 
The values of the helicity-parameter $\alpha$ are summarized in Table~\ref{Tab:heli-target}.
The CDF helicity-angle definition was also used, where 
the quantisation axis was defined as the $J/\psi$ direction of flight in the ZEUS coordinate 
system; this frame is known as the ``helicity basis''\cite{prl:85:2886}. The parameter 
$\alpha$ measured in this frame is shown in Fig.~\ref{fig-helitarg}c) and d) and listed in 
Table~\ref{Tab:heli-helbas}. The values 
$\alpha = -1$ and $\alpha = +1$ correspond to fully longitudinal and transverse polarisation, 
respectively. Within the large experimental uncertainties, the data are consistent with a 
trend from transverse to longitudinal polarisation with increasing $p_T$.

These results are consistent with the previous ZEUS measurements~\cite{zfp:c76:599}, but 
have an improved precision and extend to higher $p_T^2$. Recently, the H1 collaboration has 
also published a measurement of inelastic $J/\psi$ production~\cite{epj:c25:25} 
showing similar features to those presented here.

\subsubsection{Comparison with leading--order Colour Singlet calculations}
\label{sec:compcslo}

Figure~\ref{fig-dsdpt2} shows a comparison of the KZSZ (LO, CS) predictions with the data. 
For \mbox{$p_T^2 \sim 1~{\rm GeV^2}$}, the prediction underestimates the data by a factor of 
about two, although this is within the range of the theoretical uncertainties. For higher 
$p_T$ values, the calculation falls increasingly below the data. At 
$p_T^2 \sim 20~{\rm GeV^2}$, the LO CS prediction undershoots the data by a factor of 
$\sim 20$. 

The prediction of BKV (LO, CS) for the helicity-parameter $\alpha$ as a function of 
$p_T$ in the target frame is shown in Figs.~\ref{fig-helitarg}a) and b). The prediction 
lies somewhat below the data at low $p_T$ and somewhat above at high $p_T$, although the 
data have large statistical uncertainties. This general trend appears for both 
\mbox{0.4 $< z <$ 1} and \mbox{0.4 $< z <$ 0.9} regions. The prediction from Baranov also 
lies below the data at low $p_T$, but gives a good description of the data for 
\mbox{$p_T > 1.6$~GeV}.

The LO CS prediction from Baranov is also shown in Fig.~\ref{fig-helitarg}c) and d) 
compared to the data for the helicity-base frame. The GRV prediction was obtained by 
adding~\cite{hep-ph-9506403} a $k_T$ dependence to the GRV (collinear) gluon density. The 
KMS prediction was obtained from the $k_T$-unintegrated gluon density~\cite{pr:d56:3991}.
The data is reasonably well described by both predictions.

\subsubsection{Comparison with the next--to--leading--order Colour Singlet calculation}
\label{sec:compcsnlo}

The KZSZ (NLO, CS) prediction is compared to the data for $d\sigma/dp_T^2$ in \fig{dsdpt2}. 
The prediction is not reliable in the $p_T \rightarrow 0$ limit\cite{pl:b348:657,*np:b459:3} 
and hence is only shown for \mbox{$p_T > 1$ GeV}. The predicted shape, controlled by QCD 
gluon radiation, is in good agreement with the data. The normalisation of the predicted cross 
sections is sensitive to the assumed values of the non-perturbative QCD input parameters, 
such as the mass of the charm quark and the value of $\Lambda_{\rm QCD}$. The uncertainty 
resulting from the variation of the renormalisation and factorisation scales is small by 
comparison~\cite{priv:kraemer:2002}. The overall normalisation of the data is well 
described by the calculation, although the prediction suffers from large uncertainties. 
The discrepancy between the data and the LO CS prediction can therefore be explained by large 
NLO corrections. 

The prediction is also compared to the cross-section as a function of $W$ in 
Fig.~\ref{fig-dsdw}. The prediction again suffers from large theoretical uncertainties but 
describes the shape and normalisation of the data. Similar conclusions can be drawn from the 
comparison with the differential cross-sections $d\sigma/dz$ shown in 
Figs.~\ref{fig-dsdzpt1}--\ref{fig-dsdzpt2}. 

\subsubsection{Comparison with non--relativistic Quantum Chromodynamics calculations}
\label{sec:compcsco}

\begin{figure}
\epsfig{file=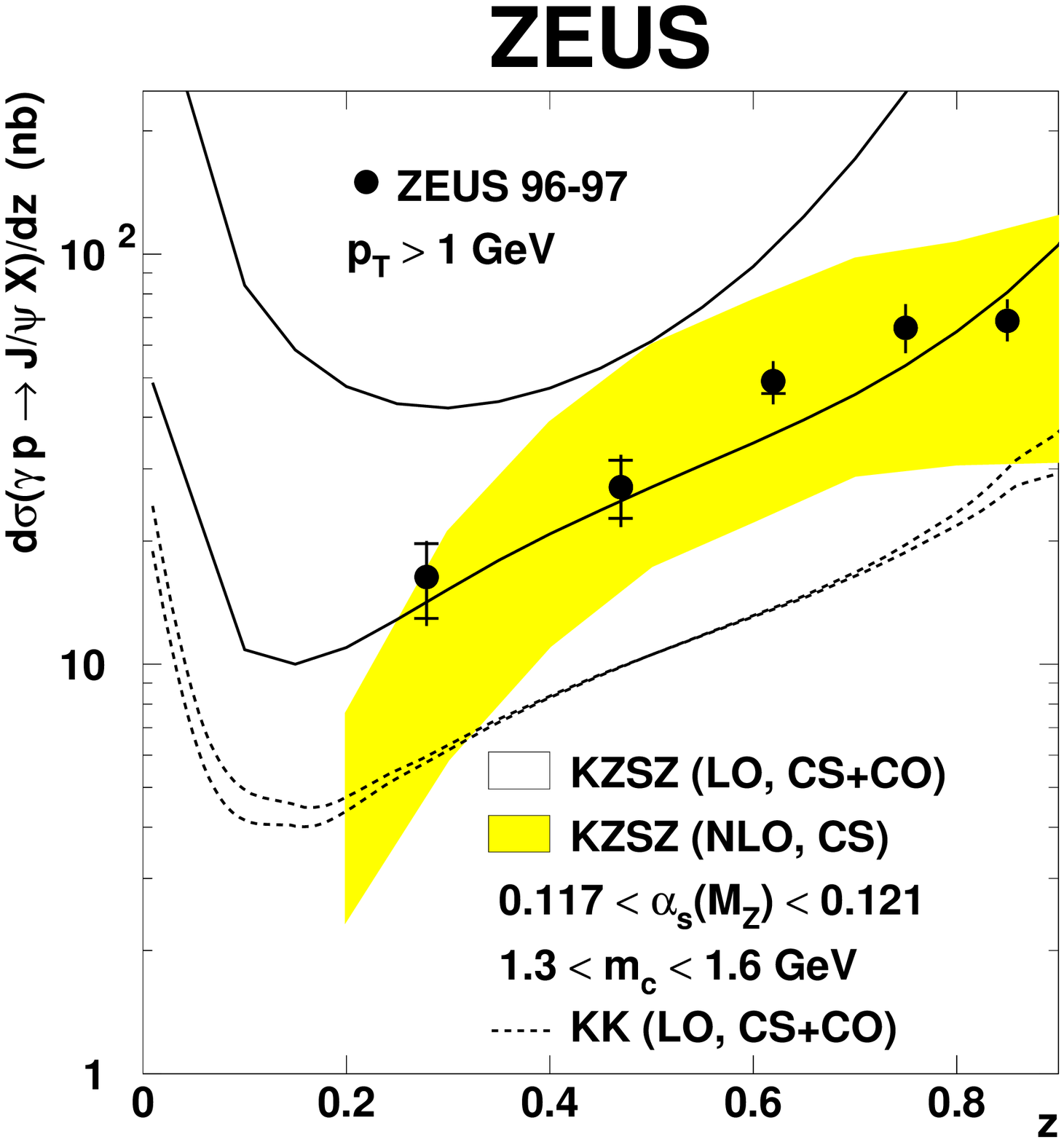,
height=15cm,width=15cm,clip=,angle=0,silent=}
\vspace{-0.5cm}
\caption{Differential cross-section $d\sigma/dz$ for $50 < W < 180$ GeV and 
\mbox{$p_T > 1$ GeV} (points). The inner error bars show the statistical uncertainty; 
the outer bars show the statistical and systematic uncertainties added in quadrature.
The shaded band shows the prediction \mbox{KZSZ (NLO, CS)}, including only the direct 
photon process. The spread in the prediction is due to uncertainties on the 
charm-quark mass and on the QCD scale parameter, $\Lambda_{QCD}$.
The solid lines show the prediction of the \mbox{KZSZ (LO, CS+CO)} calculation performed 
including both direct and resolved photon processes. 
The spread in the predictions is due to theoretical 
uncertainties in the extraction of the CO matrix elements. The KK (LO, CS+CO) 
prediction is also shown as the dashed line. The spread in the predictions is due to 
uncertainties in the extraction of the CO matrix elements.
A 15 \% contribution has been added to the predictions to account for $J/\psi$ mesons 
originating from $\psi^{\prime}$ cascade decays.}
\label{fig-dsdzpt1}
\end{figure}

\begin{figure}
\epsfig{file=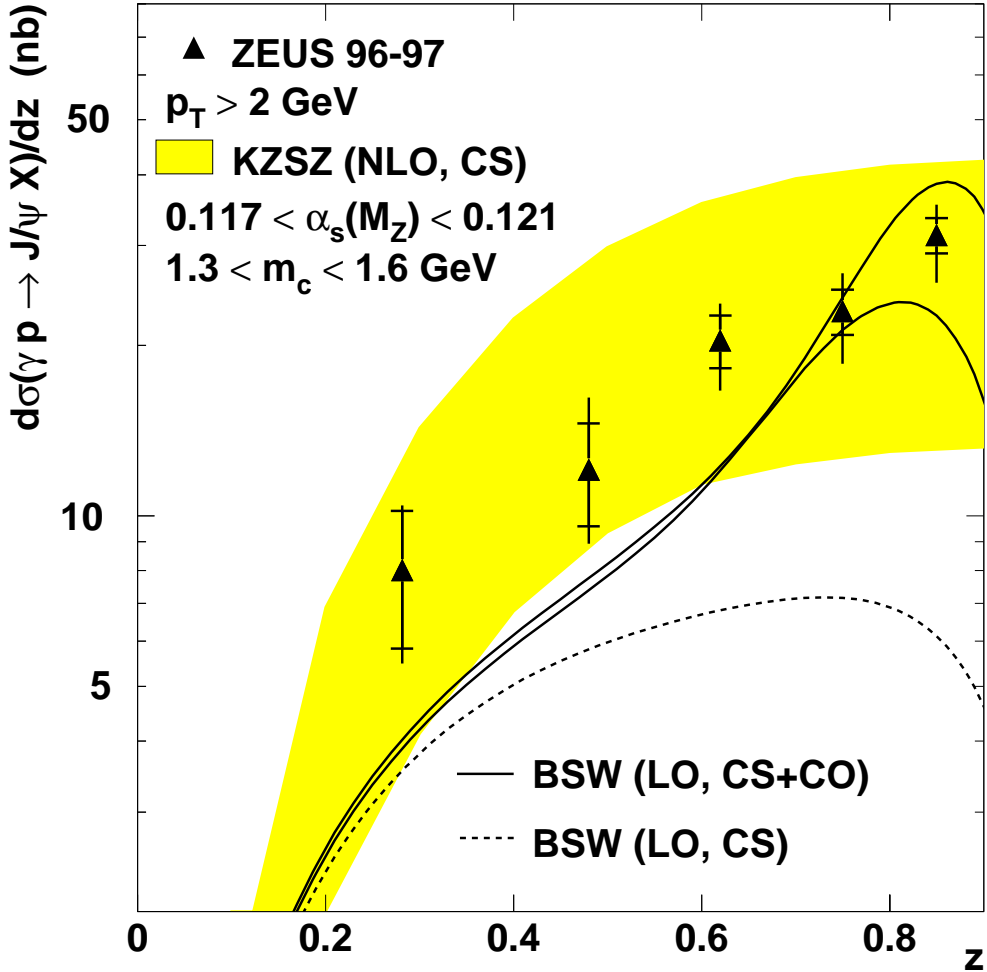,
height=15cm,width=15cm,clip=,angle=0,silent=}
\vspace{-0.5cm}
\caption{Differential cross-section $d\sigma/dz$ for $50 < W < 180$ GeV and \mbox{$p_T > 2$ GeV}. 
The inner error bars show the statistical uncertainty; the outer bars show the statistical 
and systematic uncertainties added in quadrature. The shaded band shows the prediction 
\mbox{KZSZ (NLO, CS)}. The spread in the prediction is due to 
uncertainties on the charm-quark mass and on the QCD scale parameter, $\Lambda_{QCD}$. The 
solid lines show the prediction of BSW (LO, CS+CO), where the spread in the prediction is 
due to the uncertainty on the value of the shape-function parameter. The dashed line shows 
the contribution of the CS terms only.
A 15 \% contribution has been added to the predictions to account for $J/\psi$ mesons 
originating from $\psi^{\prime}$ cascade decays.}
\label{fig-dsdzpt2}
\end{figure}

The inelasticity distributions in Figs.~\ref{fig-dsdzpt1}--\ref{fig-dsdzpt2} are compared 
with different predictions including CO matrix elements extracted by fitting independent 
data sets. In \fig{dsdzpt1}, $d\sigma/dz$ for \mbox{$p_T>1$ GeV} is compared with the 
\mbox{KZSZ (LO, CS+CO)} and \mbox{KK (LO, CS+CO)} calculations. The rise of the predicted 
cross section for $z <$ 0.1 is due to resolved photon processes. Within the large 
theoretical uncertainties, the prediction \mbox{KZSZ (LO, CS+CO)} gives a good description 
of the data. The \mbox{KK (LO, CS+CO)} result lies significantly below the data, but 
describes the shape reasonably well.

In Fig.~\ref{fig-dsdzpt2}, the differential cross-section $d\sigma/dz$ for $p_T >2$ GeV is 
compared with the \mbox{BSW (LO, CS)} and \mbox{BSW (LO, CS+CO)} 
calculations\cite{pr:d62:34004}. The CS prediction clearly lies below the data, whilst the 
inclusion of the CO terms gives a better description. The spread in the prediction, which 
is largest at high $z$, is due to the uncertainty on the value of the shape-function 
parameter. The overall shape of the spectrum is weakly dependent on the CO matrix elements, 
which primarily affect the global normalisation of the spectrum.

The KK (LO, CS+CO) prediction is compared to the cross-section $d\sigma/dy$ in \fig{dsdy} 
for the kinematic region \mbox{50 $< W <$ 180 GeV}, 0.4 $< z <$ 0.9 and $p_T >$ 1~GeV. The 
calculation includes both direct and resolved photon processes, but the resolved photon 
contribution is negligible in the selected phase-space region (due to the lower $z$ 
cut). The predicted cross section falls well below the data, although the shape is 
reasonably reproduced, as shown when the prediction is multiplied by a factor of three.

In Figs.~\ref{fig-helitarg}a) and b), the BKV (LO, CS+CO) prediction is compared to the data 
for the helicity-parameter distribution as a function of $p_T$. The spread in the prediction 
is due to theoretical uncertainties on the values of the CO matrix elements. In the 
currently accessible $p_T$ range, the CS plus CO predictions are similar to those of the CS 
only, although the prediction from the CS model rises with $p_T$, while the CS plus CO 
prediction decreases slightly.


\section{Conclusions}

Cross sections of inelastic $J/\psi$ photoproduction have been measured and 
compared with LO and NLO QCD predictions. The LO CS prediction does not describe the $p_T^2$ 
spectrum. A NLO QCD calculation in the framework of the CS model, including only the direct 
photon process, gives a good description of the $p_T^2$ and $z$ differential cross 
sections and of the cross section as a function of $W$. 
However, given the large theoretical uncertainties affecting the NLO calculation, 
it is currently not possible to constrain the size of the CO contributions. Furthermore, LO 
calculations including CO contributions, as determined from $p\bar{p}$ data, describe the 
data, albeit with large theoretical uncertainties. Although the helicity 
distribution at high $p_T$ is sensitive to the underlying production mechanism, the data 
are unable to distinguish between the two mechanisms.
These results agree with the measurements recently published by the H1 collaboration after 
taking into account, by MC extrapolation, the small ($\lesssim$ 10 \%) normalisation differences 
due to the different phase--space regions probed by the two experiments.


\section*{Acknowledgements}
 
The design, construction and installation of the ZEUS detector have been made possible by 
the ingenuity and dedicated efforts of many people from inside DESY and from the home 
institutes who are not listed as authors. Their contributions are acknowledged with great 
appreciation. The experiment was made possible by the inventiveness and the diligent 
efforts of the HERA machine group. The strong support and encouragement of the DESY 
directorate have been invaluable. We are grateful to M.~Kr\"{a}mer for providing his 
theoretical curves and for many useful discussions. We want to thank B.~A.~Kniehl, G.~Kramer, 
S.~P.~Baranov, M.~Cacciari, M.~Beneke, G.A.~Schuler, S.~Wolf and M.~Vanttinen for
making their theoretical calculations available to us.

\vfill\eject


{
\def\bibname{\Large\bf References}
\def\refname{\Large\bf References}
\pagestyle{plain}
\ifzeusbst
  \bibliographystyle{./BiBTeX/bst/l4z_default}
\fi
\ifzdrftbst
  \bibliographystyle{./BiBTeX/bst/l4z_draft}
\fi
\ifzbstepj
  \bibliographystyle{./BiBTeX/bst/l4z_epj}
\fi
\ifzbstnp
  \bibliographystyle{./BiBTeX/bst/l4z_np}
\fi
\ifzbstpl
  \bibliographystyle{./BiBTeX/bst/l4z_pl}
\fi
{\raggedright
\bibliography{./BiBTeX/user/syn.bib,%
              ./BiBTeX/bib/l4z_articles.bib,%
              ./BiBTeX/bib/l4z_books.bib,%
              ./BiBTeX/bib/l4z_conferences.bib,%
              ./BiBTeX/bib/l4z_h1.bib,%
              ./BiBTeX/bib/l4z_misc.bib,%
              ./BiBTeX/bib/l4z_old.bib,%
              ./BiBTeX/bib/l4z_preprints.bib,%
              ./BiBTeX/bib/l4z_replaced.bib,%
              ./BiBTeX/bib/l4z_temporary.bib,%
              ./BiBTeX/bib/l4z_zeus.bib}}

\providecommand{\etal}{et al.\xspace}
\providecommand{\coll}{Collab.\xspace}
\catcode`\@=11
\def\@bibitem#1{%
\ifmc@bstsupport
  \mc@iftail{#1}%
    {;\newline\ignorespaces}%
    {\ifmc@first\else.\fi\orig@bibitem{#1}}
  \mc@firstfalse
\else
  \mc@iftail{#1}%
    {\ignorespaces}%
    {\orig@bibitem{#1}}%
\fi}%
\catcode`\@=12
\begin{mcbibliography}{10}

\bibitem{pr:d23:1521}
E.~L.~Berger and D.~Jones,
\newblock Phys.\ Rev.{} {\bf D~23},~1521~(1981)\relax
\relax
\bibitem{pl:b348:657}
M.~Kr\"{a}mer \etal,
\newblock Phys.\ Lett.{} {\bf B~348},~657~(1995)\relax
\relax
\bibitem{np:b459:3}
M.~Kr\"{a}mer,
\newblock Nucl.\ Phys.{} {\bf B~459},~3~(1996)\relax
\relax
\bibitem{hep-ph-0106120}
M.~Kr\"{a}mer, Prog. Part. Nucl. Phys., {\bf 47}, 141 (2001)\relax
\relax
\bibitem{prl:76:4128}
M.~Cacciari and M.~Kr\"{a}mer,
\newblock Phys.\ Rev.\ Lett.{} {\bf 76},~4128~(1996)\relax
\relax
\bibitem{pr:d54:4312}
P.~Ko, J.~Lee and H.~S.~Song,
\newblock Phys.\ Rev.{} {\bf D~54},~4312~(1996)\relax
\relax
\bibitem{pr:d60:119902}
P.~Ko, J.~Lee and H.~S.~Song,
\newblock Phys.\ Rev.{} {\bf D~60},~119902~(1999)\relax
\relax
\bibitem{pr:d62:34004}
M.~Beneke, G.~A.~Schuler and S.~Wolf,
\newblock Phys.\ Rev.{} {\bf D~62},~34004~(2000)\relax
\relax
\bibitem{epj:c6:493}
B.~A.~Kniehl and G. Kramer,
\newblock Eur.\ Phys.\ J.{} {\bf C~6},~493~(1999)\relax
\relax
\bibitem{pr:d57:4258}
M.~Beneke, M.~Kr\"{a}mer and M.~Vanttinen,
\newblock Phys.\ Rev.{} {\bf D~57},~4258~(1998)\relax
\relax
\bibitem{pl:b428:377}
S.~P.~Baranov,
\newblock Phys.\ Lett.{} {\bf B~428},~377~(1998)\relax
\relax
\bibitem{prl:79:572}
CDF \coll, F.~Abe \etal,
\newblock Phys.\ Rev.\ Lett.{} {\bf 79},~572~(1997)\relax
\relax
\bibitem{prl:79:578}
CDF \coll, F.~Abe \etal,
\newblock Phys.\ Rev.\ Lett.{} {\bf 79},~578~(1997)\relax
\relax
\bibitem{prl:85:2886}
CDF \coll, T.~Affolder \etal,
\newblock Phys.\ Rev.\ Lett.{} {\bf 85},~2886~(2000)\relax
\relax
\bibitem{zfp:c76:599}
ZEUS \coll, J.~Breitweg \etal,
\newblock Z.\ Phys.{} {\bf C~76},~599~(1997)\relax
\relax
\bibitem{np:b472:3}
H1 \coll, S.~Aid \etal,
\newblock Nucl.\ Phys.{} {\bf B~472},~3~(1996)\relax
\relax
\bibitem{epj:c25:25}
H1 \coll, C.~Adloff \etal,
\newblock Eur.\ Phys.\ J.{} {\bf C~25},~25~(2002)\relax
\relax
\bibitem{zfp:c67:433}
M.~Gl\"uck, E.~Reya and A.~Vogt,
\newblock Z.\ Phys.{} {\bf C~67},~433~(1995)\relax
\relax
\bibitem{pr:d45:3986}
M.~Gl\"uck, E.~Reya and A.~Vogt,
\newblock Phys.\ Rev.{} {\bf D~45},~3986~(1992)\relax
\relax
\bibitem{pr:d46:1973}
M.~Gl\"uck, E.~Reya and A.~Vogt,
\newblock Phys.\ Rev.{} {\bf D~46},~1973~(1992)\relax
\relax
\bibitem{jetp:45:199}
E.A.~Kuraev, L.N.~Lipatov and V.S.~Fadin,
\newblock Sov.\ Phys.\ JETP{} {\bf 45},~199~(1977)\relax
\relax
\bibitem{sovjnp:28:822}
Ya.Ya.~Balitski\u i and L.N.~Lipatov,
\newblock Sov.\ J.\ Nucl.\ Phys.{} {\bf 28},~822~(1978)\relax
\relax
\bibitem{epj:c23:73}
A.~D.~Martin \etal,
\newblock Eur.\ Phys.\ J.{} {\bf C~23},~73~(2002)\relax
\relax
\bibitem{pr:d52:2661}
CLEO \coll, R.~Balest \etal,
\newblock Phys.\ Rev.{} {\bf D~52},~2661~(1995)\relax
\relax
\bibitem{pr:d55:1280}
H.L.~Lai \etal,
\newblock Phys.\ Rev.{} {\bf D~55},~1280~(1997)\relax
\relax
\bibitem{zeus:1993:bluebook}
ZEUS \coll, U.~Holm~(ed.),
\newblock {\em The {ZEUS} Detector}.
\newblock Status Report (unpublished), DESY, 1993,
\newblock available on
  \texttt{http://www-zeus.desy.de/bluebook/bluebook.html}\relax
\relax
\bibitem{pl:b293:465}
ZEUS \coll, M.~Derrick \etal,
\newblock Phys.\ Lett.{} {\bf B~293},~465~(1992)\relax
\relax
\bibitem{nim:a279:290}
N.~Harnew \etal,
\newblock Nucl.\ Instr.\ and Meth.{} {\bf A~279},~290~(1989)\relax
\relax
\bibitem{npps:b32:181}
B.~Foster \etal,
\newblock Nucl.\ Phys.\ Proc.\ Suppl.{} {\bf B~32},~181~(1993)\relax
\relax
\bibitem{nim:a338:254}
B.~Foster \etal,
\newblock Nucl.\ Instr.\ and Meth.{} {\bf A~338},~254~(1994)\relax
\relax
\bibitem{nim:a309:77}
M.~Derrick \etal,
\newblock Nucl.\ Instr.\ and Meth.{} {\bf A~309},~77~(1991)\relax
\relax
\bibitem{nim:a309:101}
A.~Andresen \etal,
\newblock Nucl.\ Instr.\ and Meth.{} {\bf A~309},~101~(1991)\relax
\relax
\bibitem{nim:a321:356}
A.~Caldwell \etal,
\newblock Nucl.\ Instr.\ and Meth.{} {\bf A~321},~356~(1992)\relax
\relax
\bibitem{nim:a336:23}
A.~Bernstein \etal,
\newblock Nucl.\ Instr.\ and Meth.{} {\bf A~336},~23~(1993)\relax
\relax
\bibitem{nim:a333:342}
G.~Abbiendi \etal,
\newblock Nucl.\ Instr.\ and Meth.{} {\bf A~333},~342~(1993)\relax
\relax
\bibitem{desy-92-066}
J.~Andruszk\'ow \etal,
\newblock Report \mbox{DESY-92-066}, DESY, 1992\relax
\relax
\bibitem{acpp:b32:2025}
J.~Andruszk\'ow \etal,
\newblock Acta Phys.\ Pol.{} {\bf B~32},~2025~(2001)\relax
\relax
\bibitem{nim:a365:508}
H.~Abramowicz, A.~Caldwell and R.~Sinkus,
\newblock Nucl.\ Instr.\ and Meth.{} {\bf A~365},~508~(1995)\relax
\relax
\bibitem{cpc:67:465}
G.~Marchesini \etal,
\newblock Comp.\ Phys.\ Comm.{} {\bf 67},~465~(1992)\relax
\relax
\bibitem{np:b238:492}
B.~R.~Webber,
\newblock Nucl.\ Phys.{} {\bf B~238},~492~(1984)\relax
\relax
\bibitem{cpc:43:367}
T.~Sj\"ostrand and M.~Bengtsson,
\newblock Comp.\ Phys.\ Comm.{} {\bf 43},~367~(1987)\relax
\relax
\bibitem{cpc:46:43}
H.-U.~Bengtsson and T.~Sj\"ostrand,
\newblock Comp.\ Phys.\ Comm.{} {\bf 46},~43~(1987)\relax
\relax
\bibitem{thesis:kasprzak:1994}
M.~Kasprzak,
\newblock Ph.D.\ Thesis, Warsaw University, Warsaw, Poland, Report \mbox{DESY
  F35D-96-16}, DESY, 1996\relax
\relax
\bibitem{thesis:adamczyk:1999}
L.~Adamczyk,
\newblock Ph.D.\ Thesis, University of Mining and Metallurgy, Cracow, Poland,
  Report \mbox{DESY-THESIS-1999-045}, DESY, 1999\relax
\relax
\bibitem{pr:d66:10001}
Particle Data Group, K.~Hagiwara \etal,
\newblock Phys.\ Rev.{} {\bf D~66},~10001~(2002)\relax
\relax
\bibitem{pl:b408:373}
M.~Beneke, I.~Z.~Rothstein and M.~B.~Wise,
\newblock Phys.\ Lett.{} {\bf B~408},~373~(1997)\relax
\relax
\bibitem{priv:kraemer:2002}
M. Kr\"{a}mer, private communication\relax
\relax
\bibitem{hep-ph-9506403}
J. Bl\"umlein,
\newblock Preprint \mbox{DESY-95-121} (\mbox{hep-ph/9506403}), DESY, 1995\relax
\relax
\bibitem{pr:d56:3991}
J.~Kwiecinski, A.~Martin and A.~Stasto,
\newblock Phys.\ Rev.{} {\bf D~56},~3991~(1997)\relax
\relax
\end{mcbibliography}
}
\vfill\eject


\begin{table}
\begin{center}
\begin{tabular}{|c|c|c|} \hline
$p_T$ & $<p_{T}>$ & $\sigma(\psi') / \sigma(J/\psi)$ \\
{\small(GeV)} & {\small(GeV)} &   \\ \hline\hline
 0.0~ - ~1.0~ & 0.5~~  &  0.17$\pm 0.12$ \\ \hline
 1.0~ - ~1.75 & 1.38   &  0.35$\pm 0.17$ \\ \hline
 1.75~ - ~5.0   & 3.38   &  0.26$\pm 0.15$ \\ \hline\hline
$W$ & $<W>$   & $\sigma(\psi') / \sigma(J/\psi)$ \\ 
{\small(GeV)} & {\small(GeV)} &  \\ \hline\hline
 50. -   85.  &  ~~67.5 &  0.28$\pm 0.10$ \\ \hline
 ~85. - 110.  &  ~~97.5 &  0.26$\pm 0.14$ \\ \hline
 110. - 180.  & 145.    &  0.32$\pm 0.20$ \\ \hline\hline
$z$ & $<z>$   & $\sigma(\psi') / \sigma(J/\psi)$ \\
    &         &    \\ \hline\hline
 0.55 - 0.7   & ~0.625  &  0.39$\pm 0.26$ \\ \hline
 0.7~ - 0.8   & 0.75    &  0.19$\pm 0.12$ \\ \hline
 0.8~ - 0.9   & 0.85    &  0.30$\pm 0.10$ \\ \hline\hline
\end{tabular}
\caption{Cross-section ratios between $\psi'$ and $J/\psi$ 
as function of $p_T$, $W$ and $z$ variables. These ratios are measured in the
kinematical region $50 < W < 180$ GeV and $0.55 < z < 0.9$. The 
uncertainties are statistical only.}
\label{Tab:psipratio}
\end{center}
\end{table}

\begin{table}
\begin{center}
\begin{tabular}{|c|c|c|c|} \hline
$z$ & $<z>$ & bkg & $d\sigma/dz$ \\ 
    &       & \%  & {\small(nb)} \\ \hline\hline
0.10--0.40 & 0.28 & ~0. & 26.6$\pm 4.2^{+2.7}_{-3.2}$ \\ \hline
0.40--0.55 & 0.47 & ~0. & 45.3$\pm 5.5^{+5.0}_{-4.5}$ \\ \hline
0.55--0.70 & 0.62 & ~3. & 76.3$\pm 4.1^{+8.4}_{-6.9}$ \\ \hline
0.70--0.80 & 0.75 & ~9. & ~96.7$\pm 4.4^{+13.5}_{-8.7}$ \\ \hline
0.80--0.90 & 0.85 & 34. & ~97.6$\pm 3.5^{+13.7}_{-7.8}$ \\ \hline\hline
\end{tabular}
\caption{Differential cross-section $d\sigma/dz$ measured for $50 < W < 180$ GeV
and no $p_T$ cut. In the quoted cross sections, the first uncertainty
 is statistical
and the second is systematic. Overall normalization uncertatinties due to the 
luminosity measurement ($\pm 1.6\%$) and to the $J/\psi$ decay branching ratio 
($\pm 1.7\%$) are not included in the systematic error. The column labelled 
bkg gives, in each bin, the percentage of diffractive background 
subtracted from the data.}
\label{Tab:dsdzpt0}
\end{center}
\end{table}

\begin{table}
\begin{center}
\begin{tabular}{|c|c|c|c|} \hline
$z$ & $<z>$ & bkg & $d\sigma/dz$ \\ 
    &       & \%  & {\small(nb)} \\ \hline\hline
0.10--0.40 & 0.28 & ~0. & 16.3$\pm 3.4^{+1.5}_{-2.0}$ \\ \hline
0.40--0.55 & 0.47 & ~0. & 27.1$\pm 4.4^{+3.0}_{-3.2}$ \\ \hline
0.55--0.70 & 0.62 & ~2. & 49.0$\pm 3.3^{+4.9}_{-4.9}$ \\ \hline
0.70--0.80 & 0.75 & ~8. & 66.2$\pm 3.7^{+8.6}_{-7.9}$ \\ \hline
0.80--0.90 & 0.85 & 28. & 68.9$\pm 3.1^{+8.3}_{-6.9}$ \\ \hline\hline
\end{tabular}
\caption{Differential cross-section $d\sigma/dz$ measured for $50 < W < 180$ GeV
and $p_T > 1$ GeV. In the quoted cross sections, the first uncertainty is statistical
and the second is systematic. Overall normalization uncertatinties due to the 
luminosity measurement ($\pm 1.6\%$) and to the $J/\psi$ decay branching ratio 
($\pm 1.7\%$) are not included in the systematic error. The column labelled 
bkg gives, in each bin, the percentage of diffractive background 
subtracted from the data.}
\label{Tab:dsdzpt1}
\end{center}
\end{table}

\begin{table}
\begin{center}
\begin{tabular}{|c|c|c|c|} \hline
$z$ & $<z>$ & bkg & $d\sigma/dz$ \\ 
    &       & \%  & {\small(nb)} \\ \hline\hline
0.10--0.40 & 0.28 & ~0. & ~8.0$\pm 2.2^{+1.0}_{-1.3}$ \\ \hline
0.40--0.55 & 0.48 & ~0. & 12.1$\pm 2.5^{+3.3}_{-1.9}$ \\ \hline
0.55--0.70 & 0.62 & ~0. & 20.4$\pm 2.2^{+2.4}_{-3.1}$ \\ \hline
0.70--0.80 & 0.75 & ~6. & 23.0$\pm 2.1^{+3.2}_{-3.9}$ \\ \hline
0.80--0.90 & 0.85 & 16. & 31.3$\pm 2.2^{+3.4}_{-5.0}$ \\ \hline\hline
\end{tabular}
\caption{Differential cross-section $d\sigma/dz$ measured for $50 < W < 180$ GeV
and $p_T > 2$ GeV. In the quoted cross sections, the first uncertainty
 is statistical
and the second is systematic. Overall normalization uncertatinties due to the 
luminosity measurement ($\pm 1.6\%$) and to the $J/\psi$ decay branching ratio 
($\pm 1.7\%$) are not included in the systematic error. The column labelled 
bkg gives, in each bin, the percentage of diffractive background 
subtracted from the data.}
\label{Tab:dsdzpt2}
\end{center}
\end{table}

\begin{table}
\begin{center}
\begin{tabular}{|c|c|c|c|} \hline
$p_T^2$ & $<p_T^2>$ & bkg & $d\sigma/dp_T^2$ \\ 
{\small(GeV$^2$)} & {\small(GeV$^2$)} & \% &{\small(nb/GeV$^2$)} \\ \hline\hline
 0.~~   --  ~1.   & 0.46 & 22. & 12.3$\pm 0.5 ^{+2.0}_{-0.9}$ \\ \hline
 1.~~   --  ~2.   & 1.45 & 18. & ~7.5$\pm 0.4 ^{+0.9}_{-0.5}$ \\ \hline
 2.~~   --  ~3.   & 2.45 & 16. & ~4.8$\pm 0.3 ^{+0.5}_{-0.4}$ \\ \hline
 3.~~   --  ~4.   & 3.46 & 15. & ~2.7$\pm 0.3 ^{+0.3}_{-0.3}$ \\ \hline
 4.~~   --  ~5.   & 4.47 & 14. & ~2.3$\pm 0.3 ^{+0.2}_{-0.3}$ \\ \hline
 5.~~   --  ~6.   & 5.40 & 10. & ~1.73$\pm 0.23 ^{+0.16}_{-0.24}$ \\ \hline
 6.~~   --  ~7.   & 6.45 & ~3. & ~1.56$\pm 0.21 ^{+0.14}_{-0.20}$ \\ \hline
 7.~~   --  ~8.   & 7.55 & ~0. & ~0.94$\pm 0.16 ^{+0.08}_{-0.16}$ \\ \hline
 ~8.~   ~-- ~9.5  & 8.78 & ~0. & ~0.71$\pm 0.12 ^{+0.08}_{-0.11}$ \\ \hline
 ~9.5  ~-- 11.5  & 10.48 & ~0. & ~0.50$\pm 0.08 ^{+0.05}_{-0.10}$ \\ \hline
11.5~  -- 15.5  &  13.53 & ~0. & ~0.24$\pm 0.04 ^{+0.03}_{-0.04}$ \\ \hline
15.5~  -- 21.~   & 18.02 & ~0. & ~0.15$\pm 0.02 ^{+0.04}_{-0.02}$ \\ \hline
21.~~  -- 30.~   & 24.22 & ~0. & ~~0.043$\pm 0.009 ^{+0.013}_{-0.005}$ \\ \hline\hline
\end{tabular}
\caption{Differential cross-section $d\sigma/dp_T^2$ measured for $50 < W < 180$ 
GeV and $0.4 < z < 0.9$. In the quoted cross sections, the first
uncertainty
 is statistical
and the second is systematic. Overall normalization uncertatinties due to the 
luminosity measurement ($\pm 1.6\%$) and to the $J/\psi$ decay branching ratio 
($\pm 1.7\%$) are not included in the systematic error. The column labelled 
bkg gives, in each bin, the percentage of diffractive background 
subtracted from the data.}
\label{Tab:dsdpt2}
\end{center}
\end{table}

\begin{table}
\begin{center}
\begin{tabular}{|c|c|c|c|} \hline
$W$ & $<W>$ & bkg & $\sigma$ \\ 
{\small(GeV)} & {\small(GeV)} & &{\small(nb)} \\ \hline\hline
~50~--~~70 & 61.3   & 27. &22.0$\pm 1.5 ^{+2.6}_{-2.9}$  \\ \hline
~70~--~~90 & 79.9   & 14. &24.1$\pm 1.4 ^{+2.2}_{-2.2}$  \\ \hline
~~90~--~110 & 100.1 & 10. &29.1$\pm 1.9 ^{+2.6}_{-3.5}$ \\ \hline
110~--~140 & 124.8  & ~9. &28.9$\pm 2.0 ^{+4.0}_{-3.2}$   \\ \hline
140~--~180 & 157.2  & ~4. &29.6$\pm 3.2 ^{+3.5}_{-3.0}$  \\ \hline\hline
\end{tabular}
\caption{Cross section versus $W$ measured for $0.4 < z < 0.9$
and $p_T > 1$ GeV. In the quoted cross sections, the first uncertainty
 is statistical
and the second is systematic. Overall normalization uncertatinties due to the 
luminosity measurement ($\pm 1.6\%$) and to the $J/\psi$ decay branching ratio 
($\pm 1.7\%$) are not included in the systematic error. The column labelled 
bkg gives, in each bin, the percentage of diffractive background 
subtracted from the data.}
\label{Tab:sversusW}
\end{center}
\end{table}

\begin{table}
\begin{center}
\begin{tabular}{|c|c|c|c|} \hline
$y$ & $<y>$ & bkg & $d\sigma/dy$ \\ 
    &       & \%  &{\small(nb)} \\ \hline\hline
-1.6 -- -1.2   & -1.32 & ~7. &~3.3$\pm 0.5 ^{+0.5}_{-0.4}$ \\ \hline
-1.2 -- -0.8   & -0.97 & ~9. &~7.9$\pm 0.7 ^{+0.9}_{-0.9}$ \\ \hline
-0.8 -- -0.4   & -0.59 & ~9. &11.1$\pm 0.8 ^{+1.2}_{-1.1}$  \\ \hline
-0.4~ --  0.~~~& -0.20 & 10. &10.5$\pm 0.7 ^{+1.1}_{-1.2}$  \\ \hline
 0.~ --  ~0.4  & ~0.18 & 14. &10.6$\pm 0.7 ^{+1.1}_{-1.0}$  \\ \hline
 0.4 --  ~0.8  & ~0.59 & 18. &10.9$\pm 0.7 ^{+1.2}_{-1.1}$  \\ \hline
 0.8 --  ~1.2  & ~0.92 & 17. &~8.1$\pm 0.9 ^{+1.0}_{-1.3}$ \\ \hline\hline
\end{tabular}
\caption{Differential cross-section $d\sigma/dy$ measured for $50 < W < 180$ GeV,
$0.4 < z < 0.9$ and $p_T > 1$ GeV. In the quoted cross sections, the
first uncertainty is statistical
and the second is systematic. Overall normalization uncertatinties due to the 
luminosity measurement ($\pm 1.6\%$) and to the $J/\psi$ decay branching ratio 
($\pm 1.7\%$) are not included in the systematic error. The column labelled 
bkg gives, in each bin, the percentage of diffractive background 
subtracted from the data.}
\label{Tab:dsdy}
\end{center}
\end{table}

\begin{table}
\begin{center}
\begin{tabular}{|c|c|c|c|} \hline
$p_T$ & $<p_{T}>$ & $\alpha$~~(0.4 $< z <$ 1) & $\alpha$~~(0.4 $< z <$ 0.9) \\ 
{\small(GeV)} & {\small(GeV)} &  & \\ \hline\hline
1.0 -- 1.2 & 1.10 &\ 1.12$^{+0.72}_{-0.61}$ & \ 1.12$^{+0.93}_{-0.74}$ \\ \hline
1.2 -- 1.4 & 1.30 &\ 1.02$^{+0.72}_{-0.60}$ & \ 0.82$^{+0.93}_{-0.72}$ \\ \hline
1.4 -- 1.6 & 1.50 &\ 0.76$^{+0.72}_{-0.58}$ & \ 1.44$^{+0.81}_{-0.89}$ \\ \hline
1.6 -- 1.9 & 1.74 &\ 0.32$^{+0.63}_{-0.51}$ & \ 0.30$^{+0.82}_{-0.63}$ \\ \hline
1.9 -- 2.4 & 2.13 &-0.09$^{+0.45}_{-0.38}  $ &-0.21$^{+0.53}_{-0.44}$   \\ \hline
2.4 -- 3.4 & 2.81 &-0.05$^{+0.44}_{-0.37}  $ &-0.57$^{+0.38}_{-0.32}$   \\ \hline
3.4 -- 5.0 & 4.06 &-0.24$^{+0.69}_{-0.48}  $ &-0.03$^{+0.87}_{-0.59}$   \\ \hline\hline
\end{tabular}
\caption{$J/\psi$ helicity parameter $\alpha$ as a function of $p_T$ 
measured in the target frame for $50 < W < 180$ GeV and 
$0.4 < z < 1 (0.9)$. The uncertainties are due to the total experimental 
uncertainties.}
\label{Tab:heli-target}
\end{center}
\end{table}

\begin{table}
\begin{center}
\begin{tabular}{|c|c|c|c|} \hline
$p_T$ & $<p_{T}>$ & $\alpha$~~(0.4 $< z <$ 1) & $\alpha$~~(0.4 $< z <$ 0.9) \\ 
{\small(GeV)} & {\small(GeV)} &  & \\ \hline\hline
1.0 -- 1.2 & 1.10 &  ~0.24 $^{+0.40}_{-0.33}$ & \ 0.37$^{+0.52}_{-0.42}$ \\ \hline  
1.2 -- 1.4 & 1.30 & \ 0.07 $^{+0.35}_{-0.29}$ &-0.05$^{+0.42}_{-0.35}$   \\ \hline
1.4 -- 1.6 & 1.50 &-0.07 $^{+0.35}_{-0.29}$ & \ 0.55$^{+0.62}_{-0.48}$ \\ \hline
1.6 -- 1.9 & 1.74 &-0.09 $^{+0.39}_{-0.32}$ &-0.13$^{+0.47}_{-0.38}$     \\ \hline
1.9 -- 2.4 & 2.13 &-0.28 $^{+0.39}_{-0.35}$ &-0.49$^{+0.44}_{-0.38}$     \\ \hline
2.4 -- 3.4 & 2.81 &-0.27 $^{+0.52}_{-0.42}$ &-0.58$^{+0.56}_{-0.42}$     \\ \hline
3.4 -- 5.0 & 4.06 & \ 0.39 $^{+1.44}_{-0.92}$ & \ 0.74$^{+1.51}_{-1.26}$     \\ \hline\hline
\end{tabular}
\caption{$J/\psi$ helicity parameter $\alpha$ as a function of $p_T$ 
measured in the helicity basis for $50 < W < 180$ GeV and 
$0.4 < z < 1 (0.9)$. The uncertainties are due to the total experimental 
uncertainties.}
\label{Tab:heli-helbas}
\end{center}
\end{table}



%
%
\end{document}